\documentclass[12pt]{article}

\RequirePackage{etex}
\usepackage{amssymb}
\usepackage{amsfonts}
\usepackage{amsmath}
\usepackage[nohead]{geometry}
\usepackage[singlespacing]{setspace}
\usepackage[bottom]{footmisc}
\usepackage{indentfirst}
\usepackage{endnotes}
\usepackage{graphicx}%
\usepackage{rotating}
\usepackage{xcolor}
\usepackage{enumerate}
\usepackage{hyperref}
\usepackage{undertilde}
\usepackage{tikz-network}
\usepackage[english]{babel}
\usepackage{bbold}
\usepackage{multirow}
\usepackage[longnamesfirst]{natbib}
\usepackage{tikz}
\usepackage{subcaption}
\usepackage{appendix}

\usetikzlibrary{arrows,shapes,backgrounds,automata, patterns, patterns.meta, decorations.pathreplacing}
\hypersetup{
	colorlinks=true,
	linkcolor=blue,
	citecolor=blue,
	filecolor=magenta,      
	urlcolor=cyan,
}
\newtheorem{theorem}{Theorem}

\newtheorem{corollary}{Corollary}
\newtheorem{definition}{Definition}

\newtheorem{lemma}[theorem]{Lemma}
\newtheorem{proposition}{Proposition}

\newenvironment{proof}[1][Proof]{\noindent\textbf{#1.} }{\ \rule{0.5em}{0.5em}}

\newcommand{\sgn}{\operatorname{sgn}}
\makeatletter
\def\@biblabel#1{\hspace*{-\labelsep}}
\makeatother
\begin{document}
\title{Cohesion, Ideology, and Tolerance}
\author{\textit{Patrick Allmis}\thanks{Faculty of Economics \& Janeway Institute, University of Cambridge, Austin Robinson Building, Sidgwick Avenue, Cambridge CB3 9DD, United Kingdom; E-mail: pa509@cam.ac.uk. I would like to thank Sebastian Bervoets, Yann Bramoull{\' e}, Sebastiano Della Lena, Fr{\' e}d{\' e}ric Dero{\" i}an, Matt Elliott, Mathieu Faure, Garance Genicot, Christian Ghiglino, Sanjeev Goyal, Timo Hiller, Georg Kirchsteiger, Alastair Langtry, Ana Mauleon, Luca Merlino, Paolo Pin and Agnieszka Rusinowska for their great comments and the fruitful discussions. Moreover, I thank participants of the $8^{th}$ CTE, SING17, the SasCa$2022$ PhD Conference, the CES Research Group ``Networks and Games", the YETI $2022$, the Lisbon Meetings $2023$, the PET $2023$, and the LAGV $2024$ for their great suggestions and comments.}\medskip\\{\normalsize University of Cambridge}%\medskip%\\{(preliminary draft: please do not circulate)}
}
\maketitle

\sloppy%avoids the breakage of words at the end of lines, by adjusting spaces between words inside the lines

\onehalfspacing

\begin{abstract}

\noindent Agents with different ideologies often form alliances to achieve their goals. Paradoxically, ideologically similar agents are often opponents. In this paper, ideologically heterogeneous agents choose the ideological composition of their neighborhood, their tolerance, and invest into connections. The resulting weighted network describes \textit{allies}, \textit{opponents}, and \textit{strengths}. Disputes with opponents determine benefits, which increase in an agent's strength and \textit{cohesion}. Cohesive agents have fewer mutual allies with opponents. In equilibrium, the network is segregated when cohesion is effective enough and some agents tolerate ideologically distant types to oppose closer ones. Subsidizing connections dampens polarization in societies on the verge of segregation.

\end{abstract}

\strut

\textbf{Keywords: Networks, Ideology, Tolerance, Polarization, Contests} 

\strut

\textbf{JEL Classification Numbers: C72, D74, D85} 

%\pagebreak%breaks to the next page
%\doublespacing %makes space between lines to be double, use singlespacing for space 1
\onehalfspacing
\thispagestyle{empty}
\setcounter{page}{0}
\pagebreak%breaks to the next page
%
%
%
%%%%%%%%%%%%%%%%%%%%%%The model%%%%%%%%%%%%%%%%%%%%%%%%%%%%%%%%%%%%%%%%%%%%%%%
%
%

\section{Introduction}

Factions within political parties, politically active non-governmental organizations, and interest groups share the feature of serving an underlying core of values and ideas \textemdash their ideology. These ideologies, regardless of whether they stem from convictions or serve the agenda of a specific group, dictate the goals of such politically active entities. These goals are often conflicting and agents must secure support from \textit{allies} to achieve their goals at the expense of their \textit{opponents}. %In other words, these organizations are ideological and comprise of individuals who, at least to some extent, subscribe to this ideology. 
%The collection of such ideological organizations, its members, or factions describes the ideological landscape of political agents, who cooperate with and campaign against others to achieve their often conflicting goals.%\footnote{We will use the term political agents to describe all ideological organizations in the economy.}

In contrast to what naive intuition would suggest, seemingly close agents \textemdash in terms of their ideology \textemdash are often opponents \citep{boucek2009rethinking}. We tend think of the value of an alliance as independent of other alliances, however, it is sensible to assume that potential supporters (e.g., donors or voters) would perceive it negatively were many allies of an agent in fact opponents. Indeed, \cite{greene2015consequences} document the value for political parties to be perceived as \textit{cohesive} entities with little divide between its factions.\footnote{Their argument is that supporters perceive internal divide as a signal of low competence and support this with evidence of lower vote shares in elections. This interpretation is consistent with the model we will propose, however, other interpretations are possible as well.} This interdependence between the benefits created by different alliances gives rise to a complex network describing the spillovers between allies and opponents.

%However, this is hardly a mild assumption since potential supporters of a political agent (e.g., donors or voters) would typically care about whom they cooperate with. Indeed, \cite{greene2015consequences} empirically document the value of appearing cohesive for political parties in the eyes of their supporters, where parties with less internal discord appear more cohesive. 

%One may be inclined to expect alliances between political agents who are ideologically close and attribute the collapse of an alliance to ideological cleavages within \citep{gerlach2001structure}. However, external factors also shape the organization of social movements \citep{kriesi1996organizational}. An important external factor is how potential supporters, e.g., voters or donors, value the \textit{internal cohesion} of a political agent. Indeed, \cite{greene2015consequences} empirically document the value a cohesive image for political agents, since their supporters perceive internal discord as a negative signal for the competence of a party. %In particular, political agents can undermine their \textit{internal cohesion} when failing to choose their allies appropriately.

The focus of our paper is to understand how a concern for \textit{cohesion} shapes the emerging alliances in an environment where ideological agents interact. To do so, we propose a model with ideologically heterogeneous agents who choose the ideological composition of their \textit{allies} \textemdash their \textit{tolerance}. Two agents are allies when they tolerate each others' ideologies and ideological proximity reduces the burden of tolerance \citep{murphy1997tolerance}. Intuitively, interacting with ideologically dissimilar allies entails a reputation loss in the eyes of supporters. The extent of this reputation loss is driven by the most distant allies. Agents who do not tolerate each others' ideologies are opponents. %The tolerance decisions therefore induce a signed graph. %are those agents who are intolerant towards each other's ideology.

Sustaining alliances requires effort, and even more so when an agent wants to sustain many alliances at once. To model this, we let agents exert costly effort to strengthen the connections to their allies, however, agents cannot target these efforts towards any specific ally. Instead, the weight of a connection to an ally is (i) increasing in the effort of both allies, and (ii) decreasing in the efforts of other allies that either of the two agents have. %Intuitively, the intensities of connections are diluted when more allies require attention. %Agents in our model thus choose who their neighbors are but cannot direct investment into connections beyond that.

The tolerance and effort choices determine a weighted network which describes who are allies, who are opponents, as well as the intensity of connections between allies.\footnote{Since alliances also determine oppositions, the network can be interpreted as a signed graph.} We think of the intensity of a connection as the support of agents towards each other. The network fully determines all benefits in this model. In particular, agents benefit from \textit{dispute} with their opponents to capture the idea of conflicting ideologies and the incompatible goals they dictate. An agent's benefits from a dispute increase in the number and intensity of their connections relative to their opponent. We interpret this as the expected support an agent secures for potential proposals, which we simply call their \textit{strength}.\footnote{This is similar to the idea of probabilistic voting models, e.g., \cite{austen1987interest}.} We introduce a dispute specific notion of cohesion such that an agent is more cohesive in a dispute when they share fewer connections with their opponent. Cohesion increases the effectiveness of an agent in disputing against a given opponent.%\footnote{Note, both opponents may therefore receive positive benefits from dispute.} %Intuitively, mutual allies of two opponents blur the lines between them in the eyes of supporters as suggested in \cite{greene2015consequences}. 

The effectiveness of cohesion determines the structure of the equilibrium network. Generally, allies are not ideologically as similar as possible. If cohesion is sufficiently effective, a segregated equilibrium network arises where opponents share no mutual allies. In other words, the network consists of distinct cliques. Ideologically more diverse cliques require more tolerance from its members. %\footnote{This is slightly inaccurate, since the distribution of ideologies also plays a role. However, as is apparent from Proposition \ref{proposition:polarization}, the characterization of the equilibrium does not depend on a specific type distribution.} 
If cohesion is highly effective, agents are willing to be more tolerant and society consists of few cliques where at least some of them are ideologically diverse. Otherwise, the equilibrium network is an overlapping society where opponents share mutual allies and allies are ideologically closer. 

Some important insights arise from the equilibrium characterization. First, despite homophilous preferences, i.e., agents find it easier to tolerate ideologically similar ones, agents may ally themselves with ideologically distant types and dispute against those with closer ideologies to increase their cohesion. Second, securing broad support across the ideological spectrum need not make agents more successful in achieving their goals as ideological diversity among allies may decreases their cohesion. %Through the eyes of our model, it is unsurprising why movements are fragmented or why agents with seemingly similar ideologies are often bitter rivals. %In order to appear more cohesive, ideologically similar political agents may choose to be opponents since they can market themselves more effectively by campaigning against each other. Alliances and cooperation between political actors need not indicate ideological proximity but allies may in fact be ideologically farther compared to opponents.

%Cohesive opponents may in fact both derive positive benefits from dispute in our model. We may therefore be inclined to overemphasize the benefits from compromise and cooperation in many contexts.  

The degree of \textit{polarization} induced by the endogenous network is naturally of interest in our context. We think of polarization as the total \textit{dispute intensity}, comprising of the number of disputes (extensive margin) and how intensely disputes are fought (intensive margin), which is captured by the strengths of the opponents. %Higher hurdles to interactions between agents or changes in the effectiveness of cohesion would naturally alter the equilibrium network of alliances and disputes. The effect of such shocks on dispute intensity is \textit{ex ante} ambiguous, since they may push the intensive and extensive margins in opposite directions. 

%To tame the interplay between the number of disputes and the aggregate effort, we link the equilibrium efforts of agents to the number of allies they have. 
Agents with more allies exert lower efforts in equilibrium because (i) they are in fewer disputes and investing in their strengths pays off in fewer instances, and (ii) their neighbors contribute to their strength as well, thereby crowding out their own equilibrium efforts.\footnote{A similar mechanism is present in other economic contexts, for instance, fighting efforts of armed militias \citep{konig2017networks} or collective action \citep{chwe1999structure}.} 

%This dependence between equilibrium degrees and equilibrium efforts tie the intensive margin to the extensive margin. %This guides us in understanding the overall effects of changes in the economy on dispute intensity. 
To build stronger intuition on how polarization depends on the network structure, we impose a widely used parametric structure. In particular, we let the canonical Tullock contest capture the benefits from an agent's relative strength \citep{tullock1980trials} and treat the benefits from cohesion as an additively separable component.\footnote{This is a conservative assumption since we allow agents to perfectly substitute away from investing in their cohesion.}

We are interested in how the effectiveness of cohesion affects polarization, and how this may depend on the initial network structure. The effectiveness of cohesion can increase (i) directly, when supporters value cohesion more, or (ii) indirectly, when exerting effort to intensify one's ties with allies becomes costlier. In some sense, agents substitute away from investing into their strengths and become more concerned with their cohesion instead. %Note, the cost of exerting effort influences how much agents invest in their strengths and thus also how much cohesion they want to achieve. 
%Indeed, a policy maker could influence this cost by regulating the type and quantities of campaigning available to agents, thereby influencing how much cohesion political entities want to achieve.\footnote{
For instance, restricting communication between allies may erode their support for each other in critical situations or regulations on campaign spending and content would reduce the possibility of signalling potential alliances to supporters.

The effect of a direct increase in the effectiveness of cohesion depends on the initial network structure. Agents want to have fewer mutual allies with their opponents when cohesion becomes more effective increase. %Cohesion affects equilibrium efforts only implicitly through changes in the equilibrium network structure. We can then infer how dispute intensity changes from the change in the number of disputes.\footnote{We know this because of the link between equilibrium degrees and equilibrium efforts.} 
In an initially overlapping society, the network moves towards segregation, % as benefits from cohesion increase. %Agents' neighborhoods then overlap more with the neighborhoods of their allies. 
resulting in more disputes overall. %in society, which 
This also increases aggregate equilibrium efforts and %. %This results also in greater equilibrium efforts and d
dispute intensity increases in an initially overlapping society. % on the extensive and intensive margin. 

In an initially segregated network, any agent who is in more disputes due to an increase in the effectiveness of cohesion is so because others no longer tolerate their ideology. The number of disputes in the economy must therefore fall, thereby also decreasing the aggregate equilibrium effort and decreasing dispute intensity in a segregated society. %on the extensive and intensive margin. 

%Changes in the economy have drastically different effects for different initial conditions. Greater benefits from cohesion in a segregated society increases the ideological diversity within some cliques. In initially overlapping societies, the ideological diversity among allies decreases when agents strive for greater cohesion and the effects on polarization are reversed.

%The benefits from cohesion depend on several characteristics and it is not obvious how a policy maker can influence them. A more tangible intervention would be to alter the cost of exerting effort. For instance,  Higher costs of exerting efforts increases the relative benefits from cohesion and thus the optimal tolerance decision of the political agents. 

%As alluded to before, the cost of exerting effort alters the relative benefits from cohesion since this crowds out efforts and thereby reduces the benefits agents derive from their strengths. Political agents then strive for greater cohesion to dispute effectively against opponents. In a sense, agents substitute away from convincing others of their ideas towards campaigning more effectively against opponents.

Equilibrium efforts decrease in the effort cost and thus always depress polarization on the intensive margin. However, the effort cost also determines the alliances that agents form. Again, the initial network structure plays a crucial role for the effect on polarization. %However, higher effort costs always decrease equilibrium efforts and thus dispute intensity on the intensive margin. %The effect of an increase in the effort cost on polarization then depends on how agents adapt their tolerance decisions in response. 

For high initial effort costs, the relative benefits from cohesion are high to begin with and the equilibrium network is segregated. %Intuitively, agents emphasize cohesion much more than investing in their strengths and segregate to obtain greater benefits. 
Agents rely even more on cohesion when strengthening their ties to allies becomes costlier, thereby causing (some) cliques to grow more diverse. This results in fewer disputes and dispute intensity decreases. 

In an initially overlapping society, increasing the cost of effort moves the network towards segregation. The extensive and intensive margin then go in opposing directions and how strong the crowding out of efforts is relative to the increase in the number of disputes determines the overall effect on polarization. %\footnote{This stems from the concavity of the benefit function in the strength of an agent.} 
For initially low effort costs, the crowding out of efforts is larger and dominates. This effect slows down as the network approaches segregation as agents are in more disputes. Eventually, the increase in the number of disputes (extensive margin) dominates, which is the case for an intermediate effort cost, i.e., when the network is close to segregation. %Once the network is segregated, we revert to the high-cost scenario, and dispute intensity increases with the effort cost.
%The equilibrium network structure is thus informative about whether initial effort costs are ``low", ``intermediate" or ``high". %A planner could infer the effect of an intervention from the network structure. 
Then, subsidizing efforts presents a viable option to reduce polarization.

Our model also rationalizes the paradoxical appeal that many extremist candidates have on their allies. Since extreme ideologies are at the far end of the ideological spectrum, agents with extreme ideologies are in many disputes and thus increase the cohesion of their allies by a lot. %Importantly, when extremists have more allies, they contribute less to polarization themselves while their allies contribute more, ultimately increasing polarization overall. 
Attempts to influence extremists may thus have unintended effects and actually lead to more polarization.\footnote{\cite{mostagir2023should} find a trade off in the same spirit in a model of community formation on social media.} 

%Several extensions do not qualitatively affect our results. We provide more details in Section \ref{discussion}.

The seminal work of \cite{downs1957economic} conceptualizes political actors as economic entities that are responsive to incentives. Scholars have thus become concerned with the incentives of political actors who care about winning office (e.g., \cite{riker1962theory}).\footnote{See for instance \cite{austen1988elections,baron1989bargaining} for seminal contributions on voter incentives from which we abstract in our paper.} Indeed, \cite{kriesi1996organizational} emphasize the importance of the benefits for political agents when cooperating. 

What links this literature to our paper is the idea of cohesion and its role for the benefits of agents. Cohesion contributes to a political agent's \textit{issue ownership}, i.e., how competent they are perceived to be in tackling specific issues \citep{petrocik1996issue,van2004issue,belanger2008issue}. In some sense, political agents in our model have a ``brand value" to protect \citep{snyder2002informational}. 

Our model rationalizes why ideological differences sometimes lead to the collapse of a political entity (e.g., the Christian Democratic Party in Italy in the early 1990s \citep{boucek2009rethinking}) and sometimes not (e.g., the Republican Party \citep{noel2016ideological}). In a changing society, separation into more cohesive entities becomes profitable only if the initial divide in the political landscape is not too large.%\footnote{Indeed, history has told many of such tales, e.g., the social democratic movement in Germany or the current British conservatives.}

%The ideology of an agent represents their image or identity, which agents want to protect \citep{akerlof2000economics} and strengthen through the participation in ideological organizations \citep{carvalho2016identity}.

\cite{esteban1994measurement} establish a measurement of polarization in society with the leading application of the polarization among income groups. Following their notion of polarization, our model studies what drives the emergence of polarization in the context of political factions.% and employs a measure of polarization in their spirit. 

Ideologies differentiate types horizontally and thereby induce homophily \textemdash the tendency to interact predominantly with similar individuals \citep{mcpherson2001birds,currarini2009economic,currarini2016simple}. 

Interactions in our model arise from tolerance as in \cite{genicot2022tolerance}, where individuals choose their behavior to establish connections to others and want to comply to their own ``ideal behavior". Each agent tolerates some behaviors and departs from their own ideal behavior to be tolerated by others. %and compromise is deviating from one's ideal behavior to be an acceptable connection for others.\footnote{In \cite{genicot2022tolerance} and in our paper, utilities depend on the social distance of players. \cite{iijima2017social} provide a framework to study how social distance determines the relationships between agents. Social distance also determines group memberships in heterogeneous societies \citep{baccara2013homophily,baccara2016choosing} and information consumption of agents with different demands for different types of information \citep{allmis2023homophily}.} 
Agents thus shirk their type whereas, consistent with the contexts we have in mind, types are common knowledge in our model. Tolerance is thus bilateral, whereas agents never compromise for each other in \cite{genicot2022tolerance}. Moreover, cohesion introduces an interdependence between different links, which is, in contrast to friendship networks, key in our application.\footnote{\cite{iijima2017social} provide a general framework to study how social distance determines the relationships between agents. Social distance also determines the structure of groups whose members contribute the public good of their groups \citep{baccara2013homophily,baccara2016choosing}. In our model, distance between agents determines the cost of tolerance.}

Our paper follows the network formation protocol of \cite{cabrales2011social}, where generic socialization efforts determines the intensity of interactions between agents.\footnote{This protocol has been used to study, for instance, job search via friends \citep{galeotti2014endogenous}.} A key difference to their protocol is that we include a tolerance decision, i.e., players can exclude certain types from their network. Our network formation model is thus an intermediate case between the canonical \textit{pairwise stability} criterion of \cite{jackson1996strategic} (related to tolerance in our model), and one where generic efforts jointly determine the weighted links between agents as in \cite{cabrales2011social}. Due to the complexity of our network formation protocol and in line with our application, we consider bilateral deviations as in \cite{goyal2007structural} to account for deviations in tolerance and effort decisions. 

%Social distance between economic agents naturally gives rise to polarization. \cite{esteban1994measurement} provide a measurement of polarization which we use to capture dispute intensity. 

We borrow from the well established literature on contests  \citep{tullock1967welfare,tullock1980trials,hirshleifer1989conflict} and use two well established modelling tools. First, we use weighted links as a notion of strength \citep{konig2017networks}. Second, we study the endogenous emergence of a signed network. \cite{hiller2017friends} studies signed and unweighted network formation with homogeneous agents. Our model requires a weighted network to account for the different incentives for ideologically differentiated types. This provides novel insights on the role of cohesion for segregation and polarization, as well as the role of extreme ideologies for polarization.

The remainder of the paper is organized as follows. Section \ref{model} introduces the model. Section \ref{characterization} characterizes the equilibrium. Section \ref{main} presents how the characteristics of the economy shape dispute intensity. Section \ref{extremism} discusses the role of extreme ideologies. Section \ref{discussion} describes various extensions and additional results of the model. Section \ref{conclusion} concludes. All proofs are in the appendix.

\section{The model}\label{model}

\textbf{Players:} Let $N=\{1,2,...,n\}$ be the set of players, where $i$ is the typical player. To rule out trivial cases, we assume $n\geq 3$. Player $i$ is of \textit{type} $\theta_i\in[0,1]$, which represents their ideological location or simply their \textit{ideology}. One natural interpretation of $\theta_i$ would be the extent to which $i$ favors right-wing positions. Without loss of generality, we choose $\theta_0=0$ and $\theta_1=1$ as the most extreme types. %The reader may think of types as ideologies within a single party or as ideologies across the entire political spectrum where agents are affiliated to different parties. 
Ideologies are commonly observable to everyone. Moreover, no two players have precisely the same ideology. Formally, $\theta_i$ is drawn from some cumulative distribution $\Phi$ with continuous probability density function $\varphi$. The probability to draw identical types is then zero, i.e., $P(\exists i,j:\theta_i=\theta_j)=0$ for all $i,j\in N$, and the probability that player $i$ is equally distant from some players $j$ and $h$ is zero as well, i.e., $P(\exists i,j,h\in N:\vert\theta_i-\theta_j\vert=\vert\theta_i-\theta_h\vert)=0$ for all $i,j,h\in N$.\footnote{While intuitive, this assumption is also technically convenient since it limits the concern of equilibrium multiplicity. The qualitative results of this paper, however, do not depend on it.} %, i.e., the euclidean distance between a player and someone to their left never coincides with the euclidean distance to someone to their right.\footnote{This is convenient for establishing uniqueness of an equilibrium. Relaxing this assumption would not qualitatively change the predictions of this model.} 
If $\theta_i>\theta_j$, we say $i>j$. If $\max\{1-i,i\}<\max\{1-j,j\}$, $i$ is more moderate than $j$.

\textbf{Strategies:} Strategies in this model determine the interactions among players and comprise of two actions. Each player $i$ nominates an interval of \textit{tolerable types}, denoted by $\mathbf{t}_i=[\underline{t}_i,\bar{t}_i]\in [0,1]$. We refer to this decision as their \textit{tolerance}. Denote by $\mathbf{t}=(\mathbf{t}_1,\mathbf{t}_2,...,\mathbf{t}_n)$ the vector of tolerance choices. It is costly to tolerate different ideologies and even more so if these ideologies are farther from one's own ideology. For simplicity, let the total cost of tolerance associated with a given tolerance interval for player $i$ be $\tau[(\underline{t}_i-\theta_i)^2+(\bar{t}_i-\theta_i)^2]$, where the parameter $\tau>0$ captures the \textit{flexibility} or \textit{stubbornness} of agents.\footnote{Any specification that is increasing and convex in the distance between an agent's type and the bounds of the tolerance interval delivers the qualitative results presented in this paper.} For lower values of $\tau$, agents are more flexible.\footnote{In Section \ref{discussion}, we discuss the possibility of $\tau_i\ne\tau_j$.} The tolerance cost can, for instance, represent a reputation loss from interacting with individuals of a different ideology. Such a reputation loss would be substantially larger if an agent interacted with someone far from their own ideology. Moreover, a reputation loss of this form would be driven by the ``most distant ideology" one tolerates. We allow players to choose asymmetric tolerance intervals, so generally, $(\underline{t}_i-\theta_i)^2\ne(\bar{t}_i-\theta_i)^2$. In principle, players may choose to tolerate types far to the right (left) of their own ideology and simultaneously be intolerant towards more similar ideologies to their left (right).\footnote{Note, we do not consider the cases $\bar{t}_i=\underline{t}_i>\theta_i$ ($\bar{t}_i=\underline{t}_i<\theta_i$) or $\underline{t}_i<0$ ($\bar{t}_i>1$) since such strategies would obviously never occur in an equilibrium.}

Each player $i$ chooses a generic effort to establish connections to others which we denote by $x_i\in X_i=[0,+\infty)$. Effort entails a constant marginal cost $c>0$. The parameter $c$ may represent regulations on the communication between agents or their campaigning efforts, e.g., whether they can signal future alliances to supporters. The effort profile is then a vector $x=(x_1,x_2,...,x_n)$. 

We write the strategy of player $i$ as $s_i\in S_i=S\equiv X\times T$, where $\mathbf{t}_i\in T=[0,1]$. %\footnote{In principle, we need not restrict $T=[0,1]$ but do so since agents would never tolerate outside the type space anyways.} %Strategies comprise of the socialization effort and the tolerance choice. 
Denote by $s=(s_i,s_{-i})$ the strategy profile, where $s_{-i}$ denotes the strategies of all players, other than $i$. 

\textbf{Network formation:} Our network formation protocol allows players to choose the composition of their neighborhoods through their tolerance decision and the intensity of their interactions through their efforts. The two choices jointly generate a weighted network, which describes \textit{alliances}, i.e., links of strictly positive weight, and \textit{disputes}, i.e., links of weight zero. One can thus interpret the resulting network as a signed graph of positive (alliances) and antagonistic relationships (disputes). We denote this network by an adjacency matrix $g$ and, abusing notation, refer to $g$ simply as the network. Let $K_i(g)=\{j\in N:\theta_j\in[\underline{t}_i,\bar{t}_i]\}$ denote the set of players, whose types fall into $i$'s tolerance interval. Denote by $k_i(g)$ the cardinality of this set.\footnote{Note, we also use this notation to denote the neighborhood of player $i$. These coincide straightforwardly in equilibrium.} We sometimes write $K_i$ for $K_i(g)$ and $k_i$ for $k_i(g)$ when no confusion arises. Define a weighting function 

\begin{equation}
	\rho_{ij}(x,\mathbf{t})=
	\begin{cases}
		\frac{1}{\sum_{h\in K_i\cup K_j}x_h}\text{ if }j:\theta_j\in\mathbf{t}_i\text{, }\theta_i\in\mathbf{t}_j\text{, }x_i>0 \text{ and }x_j>0,\\
		0 \text{ otherwise}
	\end{cases}
\end{equation}

There is no link between $i$ and $j$ whenever $i$ does not tolerate $j$'s ideology ($\theta_j\not\in[\underline{t}_i,\bar{t}_i]$) or vice versa. Similarly, there is no link between $i$ and $j$ when either player exerts no effort to strengthen connections, i.e., $x_i=0$ or $x_j=0$. In all other cases, links are weighted by the aggregate socialization effort of every potential neighbor of $i$ or $j$ (everyone in their respective tolerance intervals). Congestion effects therefore reduce the intensity of a connection between two nodes when other %potential 
connections exert high efforts. 

We obtain the network $g$ by setting

\begin{equation}
	g_{ij}(x,\mathbf{t})=\min\{\rho_{ij}(x,\mathbf{t})x_ix_j,1\}
\end{equation}

We simply write $g_{ij}$ when doing so creates no confusion. Ultimately, we want to interpret $g_{ij}$ as the expected support for $i$ from their neighbor $j$. For this reason, we impose $g_{ij}\leq1$ for all $i,j\in N$. Connections strengthen as an agent invests more into their connections, all else equal, and even more so if their neighbors invest more into their connections as well. We capture this idea by letting the weight of a link depend on the product of the socialization efforts. 

Following this protocol, $\rho_{ij}(x,\mathbf{t})=\rho_{ji}(x,\mathbf{t})$ and links are symmetric, i.e., $g_{ij}=g_{ji}$ for all $i,j\in N$. Moreover, self-loops are a genuine feature of this model, so $g_{ii}\in [0,1]$. Self-loops allow agents to pursue their own agenda independently from others.

%The weighted network $g$ captures to whom each agent $i$ is \textit{connected}, i.e., whose types $i$ tolerates and who tolerates her. Moreover, $g$ captures how much players socialize with each of their connections. %For simplicity, agents distribute their socialization effort evenly across players whom they tolerate. %Throughout the paper, the following definitions will be of use.

%We borrow this network formation protocol from \cite{cabrales2011social}. We augment their protocol by a tolerance decision which determines the composition of an agent's neighborhood. 

\textbf{Network definitions:} A player $i$ is \textit{isolated} if $g_{ij}=0$ for all $j\in N\setminus\{i\}$. The network $g$ is empty if every agent agent is isolated.

A player $j$ is a \textit{neighbor} of $i$ whenever $g_{ij}>0$. Player $j$ is an \textit{opponent} of $i$ whenever $i$ and $j$ are not neighbors, i.e., $g_{ij}=0$. In this case, $i$ and $j$ are in \textit{dispute}. %One can interpret the network $g$ as a signed graph.

A player's \textit{degree} is the number of their connections, which we call $k_i$. %with some abuse of notation.

A subset of players is a clique $C(g)$, if for all $i\in C(g)\subseteq N$, $g_{ij}>0$ if $j\in C(g)$ and $g_{ij}=0$ otherwise. %The network $g$ is complete, if all players are in the same clique. 

A network $g$ is ordered with respect to types, if for any two nodes $i$ and $j$, with $i>j$, $\bar{t}_i\geq\bar{t}_j$ and $\underline{t}_i\geq\underline{t}_j$.

A network $g$ is \textit{segregated} if it consists of cliques. A segregated network exhibits \textit{strong structural balance} if it consists of exactly two cliques and \textit{weak structural balance} otherwise \citep{cartwright1956structural}. %\footnote{This notion exhibits similar to the notion of structural balance, i.e., }

A network is an \textit{overlapping society} if it is not segregated.

\textbf{Benefits:} The network determines all disputes and the resulting benefits from them. Connections influence a player's \textit{strength} and how effectively they campaign against their opponents. Establishing links to others yields only indirect benefits through altering the benefits from disputes. In a model of competing ideological agendas, it is natural to model dispute as a contest between two players. An agent's strength, captured by the sum of the weights of all their links, i.e., $\lambda_i(g)=\sum_{j\in N}g_{ij}$, uniformly increases their benefits in all disputes.\footnote{We can study alternative versions of this notion where higher order neighbors also contribute to an agent's strength or where common neighbors do not contribute to the strength of players in dispute (see Section \ref{discussion}).} This is in line with interpreting strengths as the expected support from allies. 

Another ingredient of our model is how effectively opponents campaign against each other. We posit that this depends on the \textit{cohesion} of an agent. An agent is more cohesive in a dispute if they share many connections with the opponents of their opponent. %Conversely, if a player interacts with many opponents of their opponent, they can differentiate more credibly from their opponent, i.e., they dispute more effectively against said opponent. 
Formally, let $\lambda_{ij}(g)=\sum_{j\in N}\sgn(g_{ih})(1-\sgn(g_{hj}))$ denote the number of $i$'s neighbors who are in dispute with $i$'s opponent $j$. The function $\sgn$ denotes the sign function, which equals one if $g_{ih}>0$ and zero otherwise. The \textit{dispute technology} thus composes of a global component (the strengths of players) and a dispute-specific component (the dispute-specific cohesion).

Denote the benefits for player $i$ from a dispute with player $j$ by $f(\lambda_i(g),\lambda_j(g),\lambda_{ij}(g))$. Call $f(\cdot,\cdot,\cdot)$ the contest success function (henceforth: CSF). For convenience, we write $\lambda_i$ for $\lambda_i(g)$ and $f(\lambda_i,\lambda_j,\lambda_{ij})$ for $f(\lambda_i(g),\lambda_j(g),\lambda_{ij}(g))$ when no confusion arises. The function $f(\cdot,\cdot,\cdot)$ is strictly increasing in its first argument, the strength of the player, and concave. Moreover, $f(\cdot,\cdot,\cdot)$ is strictly decreasing and convex in its second argument, the strength of the opponent. We normalize $f(\lambda_i,\lambda_j,\lambda_{ij})=0$ for all $\lambda_i,\lambda_j$ such that $\lambda_i=\lambda_j\text{ and }\lambda_{ij}=0$. %Our framework therefore nests the case when dispute technology only depends on agents' strengths. %In this case, players of equal strength are equally likely to win and gain benefits of zero from a dispute. 
Let $f(\cdot,\cdot,\cdot)$ be increasing in the third argument. %Parameters $\beta\geq0$ and $\alpha\geq0$ captures how efficient dispute technology is.  

Define a function $\delta(y)=f(\lambda_i,\lambda_j,y)-f(\lambda_i,\lambda_j,y-1)$ for all $y\in\{1,...,n-2\}$.\footnote{Note, $n-2$ is the maximum number of agents, an individual can have a mutual opponent with.} The function $\delta(y)$ captures the additional benefits for player $i$ from dispute with player $j$ when adding the $y^{th}$ opponent of $j$ to their neighborhood, all else equal. Define $\underline{\delta}=\min_{y\leq n-2}\delta(y)$ as the lowest value this function can possibly take. The value $\underline{\delta}$ establishes a lower bound on how effectively a player can campaign against others. %For agents in society $N$, 
We say cohesion is more effective %compared to agents in society $N'$ if 
when $\underline{\delta}$ is larger.

The utility of player $i$ is given by

\begin{equation}
u_i(s)=\sum_{j\not\in K_i(g)}f(\lambda_i(g),\lambda_j(g),\lambda_{ij}(g))-cx_i-\tau\left[(\theta_i-\underline{t}_i)^2+(\theta_i-\bar{t}_i)^2\right]%-\eta^-_i(g)k
\end{equation}

The utility of an agent comprises of the benefits from each dispute (first term), the cost of their effort (second term) and their cost of tolerance (third term).

To build stronger intuition, we will use the normalized Tullock CSF, augmented by cohesion, later on to establish comparative statics. 

\begin{definition}\label{definition:}
	The normalized CSF in ratio form is
	
		\begin{equation}
			f(\lambda_i,\lambda_j,\lambda_{ij})=\frac{\lambda^\phi_i}{\lambda^\phi_i+\lambda^\phi_j}-\frac{1}{2}+\beta \lambda_{ij}^\alpha
		\end{equation}

	and in difference form 

		\begin{equation}
			f(\lambda_i,\lambda_j,\lambda_{ij})=\frac{1}{1+e^{\phi(\lambda_j-\lambda_i)}}-\frac{1}{2}+\beta \lambda_{ij}^\alpha
		\end{equation}

\end{definition}

The values of $\phi\in[0,1]$\footnote{In principle, we could generalize the model to $\phi>1$.}, $\beta\geq0$ and $\alpha\geq0$ parameterize the CSF. Higher values of $\phi$ favor stronger agents. The parameter $\beta$ scales the value of cohesion and $\alpha$ captures the marginal changes in the value of cohesion.\footnote{While the Tullock contest poses a natural framework to think about disputes, other forms of modelling disputes are possible. We refer the reader to Appendix \ref{appendixB}.}
 
%Our interest is to find a network where no pair of players can jointly deviate to increase their utilities. 
A strategy profile $s^\ast$ is a \textit{bilateral equilibrium} (BE) \citep{goyal2007structural}, if 

\begin{enumerate}[{(}i{)}]
	\item $u_i(s^\ast_i,s^\ast_{-i})\geq u_i(s^\ast_i,s'_{-i})\;\forall i\in N \text{ and }s'\in S$;
    \item for each player pair $i,j\in N$, and strategy pair $(s_i,s_j)$, \\$u_i(s_i,s_j,s^\ast_{-i-j})>u_i(s^\ast_i,s^\ast_j,s^\ast_{-i-j})$ implies $u_j(s_i,s_j,s^\ast_{-i-j})<u_j(s^\ast_i,s^\ast_j,s^\ast_{-i-j})$.
    
\end{enumerate}

Our solution concept requires that no agent can increase their utility by changing their strategy (condition (i)) and that any conceivable bilateral deviation of agents $i$ and $j$ reduces the utility of at least on of the two agents (condition (ii)). 

Total effort is the sum of individual efforts, $\sum_{i\in N}x_i$. 

%Welfare is the sum of individual utilities, $WF(s)=\sum_{i\in N}u_i(s)$. 

We require a measure of divide in society. A natural candidate is polarization captured by the aggregate intensity of disputes \citep{esteban1994measurement}. Formally, dispute intensity is $\iota(s)=\sum_{i\in N}\sum_{j\ne i}\lambda_i(1-\sgn(g_{ij}))\lambda_j$. Dispute intensity increases on the extensive margin when there are more disputes. Dispute intensity increases on the intensive margin if players' strengths increase.% or \cite{HO2020}. %\sum_{i\in N}x_i\left(\sum_{h\not\in K_i}x_h\right)$.

%Define a \textit{dense society} such that $n\to\infty$ and $P(\exists i\in N:\theta_i\in[\theta-\epsilon,\theta+\epsilon])\to1$.

%I discuss several foundations of the modelling assumptions in Appendix \ref{appendixB}.

\section{Equilibrium characterization}\label{characterization}

We now proceed to characterizing the equilibrium of the game. We present conditions under which a bilateral equilibrium network exhibits (strong) structural balance, or is an overlapping society. First, we establish a powerful preliminary result.

\begin{lemma}\label{lemma:ordered}
	Any bilateral equilibrium network is ordered. 
\end{lemma}

According to Lemma \ref{lemma:ordered}, we can infer the ordering of the bounds of the tolerance intervals from the types. Players in our model are \textit{ex ante} identical apart from their ideologies. Ideologies only affect the costs of tolerating and being tolerated by others. %Despite incentives to interact with agents who are ideologically farther when this increases one's cohesion, %in disputes, homophilous preferences would still push agents to choose their allies to be as close as possible subject to the disputes they are in. A
Agents want to achieve a certain level of cohesion at the lowest possible tolerance cost. Orderedness arises since all agents behave this way.

Orderedness is a powerful result since it restricts the deviations we have to consider when characterizing the BE network. However, let us first note existence and uniqueness. %of the BE network.

\begin{proposition}\label{proposition:existence}
	There exists a unique bilateral equilibrium of the game.
\end{proposition}

We use a constructive argument to prove existence. Clearly, players only tolerate those who tolerate them as well. Otherwise, those players are in dispute and one agent has a profitable deviation in narrowing their tolerance interval (since Lemma \ref{lemma:ordered} establishes orderedness). We then proceed to let $1$ and $0$ choose their tolerance interval. Players $1$ and $0$ never tolerate each other, since they would be in no disputes and receive zero benefits. Either $1$, $0$, or both, would then receive a strictly greater utility from being in dispute. Since agents always bilaterally agree to tolerate each other in equilibrium, only players who tolerate $1$ ($0$) are also in $1$'s ($0$'s) tolerance interval. Among all players who are not connected to $1$ or $0$, we can take the most extreme type. If no such player exists, we reach an equilibrium by letting everyone else choose their tolerance intervals since this would not induce any profitable deviation for anyone who has already made their tolerance decision. Otherwise, we can let this player choose their tolerance interval and repeat this process until no players remain. By construction, no profitable deviation emerges. %for any player who has already chosen their tolerance interval. %All types who are more extreme than $i$ are of no concern since the BE network is ordered. We can simply iterate this process until all players have chosen their tolerance interval, which 
This process eventually yields a BE of the game, thereby proving existence. %This is because by our assumptions on the CSF, the tolerance decisions also determine the effort profile. Existence follows. 

The endogenous tolerance decisions define who are allies and opponents and thus uniquely pin down the equilibrium effort profile. Hence, if there existed multiple BE networks, there would exist multiple equilibrium tolerance profiles. At least one of two things must then be true. Either, (i) at least one player is indifferent between tolerating and not tolerating some other type, or (ii), some player is indifferent between tolerating two distinct types. In case (i), the benefits of a connection must precisely coincide with the additional cost of tolerance. Since agents derive benefits from each dispute, the sum of all benefits is not a continuous function. Given our assumptions on the type distribution, such instances have zero probability. In case (ii), establishing either of two connections yields exactly the same net benefits. Since agents always differ somewhat in their ideology, this indifference occurs with probability zero as well and the BE network is indeed unique.\footnote{While bilateral equilibrium is a natural solution concept in this context, employing weaker concepts would also be possible. However, issues of equilibrium multiplicity would arise.}

Having established existence and uniqueness, we next turn to characterizing the BE network. The dispute technology and players' ideologies jointly determine the incentives for interactions in this model. Dispute technology comprises of an agent's strength, i.e., the number of allies and the intensity of connections to them, and how cohesive the agents are in disputes, i.e., the number allies who are opponent of their opponent as well. The benefits from cohesion push towards segregation, since players want to coordinate their neighborhoods to share many opponents with their allies. Homophily pushes towards overlaps in the neighborhoods of players in dispute. The structure of the BE network therefore depends critically on how effective cohesion is, and hence $\delta(y)$. %for all possible values of $y$. 
 
The next Proposition provides a full characterization of the unique BE network subject to the effectiveness of cohesion. 

\begin{proposition}\label{proposition:structural-balance}
	There exist thresholds $\delta^{\ast}$ and $\delta^{\ast\ast}$, with $\delta^{\ast\ast}\geq\delta^\ast$, such that 
		\begin{enumerate}[{(}i{)}]
		
		    \item if the effectiveness of cohesion is high ($\underline{\delta}\geq\delta^{\ast\ast}$), the bilateral equilibrium network exhibits strong structural balance;
			
			\item if the effectiveness of cohesion is intermediate ($\delta^{\ast\ast}>\underline{\delta}\geq\delta^\ast$), the bilateral equilibrium network exhibits weak structural balance;

			\item otherwise, the network is an overlapping society.

		\end{enumerate}
\end{proposition}

Proposition \ref{proposition:structural-balance} establishes the link from cohesion to the BE network structure. It is beneficial to have similar allies as one's allies if cohesion is highly effective. Agents are then willing to tolerate even distant ideologies as long as this increases their cohesion in disputes with opponents who are possibly ideologically more similar. The resulting network comprises of two cliques, where agents are allies with everyone in their clique and opponents of everyone else. At least one of the cliques contains agents of greatly different ideologies and sustaining it requires high tolerance. This is beneficial only when cohesion generates sufficiently high benefits. For intermediate effectiveness of cohesion, agents are not willing to tolerate as far. %Since agents then choose a narrower revelation interval, c
Cliques are consequently smaller and more homogeneous and the network comprises of more than two cliques.

In all other cases, at least some players ally themselves with opponents of an ally. This requires less tolerance and players forgo some benefits from cohesion to reduce the burden of tolerating other ideologies. % We call such a network an overlapping society.

%Figures \ref{fig-propiia} and \ref{fig-propiib} illustrate examples.

\begin{figure}[ht]
    \centering
    \begin{subfigure}[b]{0.48\linewidth}
    \begin{tikzpicture}[
	contributor/.style={diamond,draw,thick, fill=blue!20, scale=0.7},
	periphery/.style={circle,draw,thick, scale=0.7, fill=yellow!20},
	core/.style={regular polygon,regular polygon sides=5,draw,thick, fill=green!20, scale=0.7},
	isolated/.style={regular polygon,regular polygon sides=6,draw,thick, fill=red!20, scale=0.7},
	]
	\path[draw,thick,below] (0,0) edge node {$\theta$} (7,0);
	\path[draw,thick,left] (0,0) edge node {$x^\ast_{i}$} (0,3.5);
	\node[periphery] at (7,2.8) (10){1};
	\node[periphery] at (0,2.8) (0){0};
	\node[periphery] at (6.25,1) (9){};
	\node[periphery] at (5.8,2) (8){};
	\node[periphery] at (5.2,1.2) (8a){};
	\node[periphery] at (4.2,1.8) (7){};
	\node[periphery] at (3.6,1.4) (6){};
	\node[periphery] at (3,1.4) (5){};
	\node[periphery] at (0.5,1) (1){};
	\node[periphery] at (1,2) (2){};
	\node[periphery] at (1.5,1.2) (3){};
	\node[periphery] at (2.5,1.8) (4){};
	
	\draw (0) to (1);
	\draw (1) to (2);
	\draw (2) to (3);
	\draw (3) to (4);
	\draw (4) to (5);
	\draw (5) to (6);
	\draw (6) to (7);
	\draw (7) to (8a);
	\draw (8a) to (8);
	\draw (8) to (9);
	\draw (9) to (10);
	
	\end{tikzpicture}
	\caption{$\underline{\delta}<\delta^\ast$}
	\end{subfigure}
	\begin{subfigure}[b]{0.5\linewidth}
    \begin{tikzpicture}[
	contributor/.style={diamond,draw,thick, fill=blue!20, scale=0.7},
	periphery/.style={circle,draw,thick, scale=0.7, fill=yellow!20},
	core/.style={regular polygon,regular polygon sides=5,draw,thick, fill=green!20, scale=0.7},
	isolated/.style={regular polygon,regular polygon sides=6,draw,thick, fill=red!20, scale=0.7},
	]
	\path[draw,thick,below] (0,0) edge node {$\theta$} (7,0);
	\path[draw,thick,left] (0,0) edge node {$x^\ast_{i}$} (0,3.5);
	\node[contributor] at (7,1.25) (10){1};
	\node[isolated] at (0,1.25) (0){0};
	\node[contributor] at (6.25,1.25) (9){};
	\node[contributor] at (5.8,1.25) (8){};
	\node[contributor] at (5.2,1.25) (8a){};
	\node[core] at (4.2,1.25) (7){};
	\node[core] at (3.6,1.25) (6){};
	\node[core] at (3,1.25) (5){};
	\node[isolated] at (0.5,1.25) (1){};
	\node[isolated] at (1,1.25) (2){};
	\node[isolated] at (1.5,1.25) (3){};
	\node[core] at (2.5,1.25) (4){};
	
	 \draw (0) to (1);
	 \draw (1) to (2);
	 \draw (2) to (3);
	 \draw (0) to[out=45, in=135] (2);
	 \draw (3) to[out=135, in=45] (1);
	 \draw (3) to[out=225, in=-45] (0);
	
	 \draw (9) to (10);
	 \draw (8) to (9);
	 \draw (8) to[out=45, in=135] (10);
	 \draw (8a) to[out=-45, in=-135] (10);
	 \draw (8a) to[out=45, in=135] (9);
	 \draw (8a) to (8);
	
	 \draw (4) to (5);
	 \draw (5) to (6);
	 \draw (6) to (7);
	 \draw (7) to[out=225, in=-45] (4);
	 \draw (7) to[out=135, in=45] (5);
	 \draw (4) to[out=45, in=135] (6);
    %(5)  [->] edge node [left] {} (2);
	
	\end{tikzpicture}
	\caption{$\underline{\delta}\in[\delta^\ast,\delta^{\ast\ast})$}
    \end{subfigure}
    \begin{tikzpicture}[
	contributor/.style={diamond,draw,thick, fill=blue!20, scale=0.7},
	periphery/.style={circle,draw,thick, scale=0.7, fill=yellow!20},
	core/.style={regular polygon,regular polygon sides=5,draw,thick, fill=green!20, scale=0.7},
	]
	%\matrix [draw,below left] at (current bounding box.north west) {
  %\node [contributor,label=right:Contributor] {}; &
  %\node [core,label=right:Core] {}; &
  %\node [periphery,label=right:Periphery] {}; \\
%};
    \end{tikzpicture}
\caption{$f(\cdot,\cdot,\cdot)=\frac{\lambda_i^\phi}{\lambda_i^\phi+\lambda_j^\phi}-\frac{1}{2}+\beta\lambda^\alpha_{ij}$, $c=1$, $\tau=1$}
\label{fig-propiia}
\end{figure}

\begin{figure}[ht]
    \centering
    \begin{subfigure}[b]{0.48\linewidth}
    \begin{tikzpicture}[
	contributor/.style={diamond,draw,thick, fill=blue!20, scale=0.7},
	periphery/.style={circle,draw,thick, scale=0.7, fill=yellow!20},
	core/.style={regular polygon,regular polygon sides=5,draw,thick, fill=green!20, scale=0.7},
	isolated/.style={regular polygon,regular polygon sides=6,draw,thick, fill=red!20, scale=0.7},
	]
	\path[draw,thick,below] (0,0) edge node {$\theta$} (7,0);
	\path[draw,thick,left] (0,0) edge node {$x^\ast_{i}$} (0,3.5);
	\node[contributor] at (7,3) (10){1};
	\node[isolated] at (0,0.5) (0){0};
	\node[isolated] at (6.25,0.5) (9){};
	\node[isolated] at (5.8,0.5) (8){};
	\node[isolated] at (5.2,0.5) (8a){};
	\node[isolated] at (4.2,0.5) (7){};
	\node[isolated] at (3.6,0.5) (6){};
	\node[isolated] at (3,0.5) (5){};
	\node[isolated] at (0.5,0.5) (1){};
	\node[isolated] at (1,0.5) (2){};
	\node[isolated] at (1.5,0.5) (3){};
	\node[isolated] at (2.5,0.5) (4){};
	
	\draw (0) to (1);
	\draw (1) to (2);
	\draw (2) to (3);
	\draw (3) to (4);
	\draw (4) to (5);
	\draw (5) to (6);
	\draw (6) to (7);
	\draw (7) to (8a);
	\draw (8a) to (8);
	\draw (8) to (9);
	\draw (0) to[out=45, in=135] (9);
	
	\end{tikzpicture}
	\caption{$\underline{\delta}>\delta^{\ast\ast}$}
	\end{subfigure}
	\begin{subfigure}[b]{0.48\linewidth}
    \begin{tikzpicture}[
	contributor/.style={diamond,draw,thick, fill=blue!20, scale=0.7},
	periphery/.style={circle,draw,thick, scale=0.7, fill=yellow!20},
	core/.style={regular polygon,regular polygon sides=5,draw,thick, fill=green!20, scale=0.7},
	isolated/.style={regular polygon,regular polygon sides=6,draw,thick, fill=red!20, scale=0.7},
	]
	\path[draw,thick,below] (0,0) edge node {$\theta$} (7,0);
	\path[draw,thick,left] (0,0) edge node {$x^\ast_{i}$} (0,3.5);
	\node[contributor] at (7,1.25) (10){1};
	\node[isolated] at (0,1.25) (0){0};
	\node[contributor] at (6.25,1.25) (9){};
	\node[contributor] at (5.8,1.25) (8){};
	\node[contributor] at (5.2,1.25) (8a){};
	\node[core] at (4.2,1.25) (7){};
	\node[core] at (3.6,1.25) (6){};
	\node[core] at (3,1.25) (5){};
	\node[isolated] at (0.5,1.25) (1){};
	\node[isolated] at (1,1.25) (2){};
	\node[isolated] at (1.5,1.25) (3){};
	\node[core] at (2.5,1.25) (4){};
	
	 \draw (0) to (1);
	 \draw (1) to (2);
	 \draw (2) to (3);
	 \draw (0) to[out=45, in=135] (2);
	 \draw (3) to[out=135, in=45] (1);
	 \draw (3) to[out=225, in=-45] (0);
	
	 \draw (9) to (10);
	 \draw (8) to (9);
	 \draw (8) to[out=45, in=135] (10);
	 \draw (8a) to[out=-45, in=-135] (10);
	 \draw (8a) to[out=45, in=135] (9);
	 \draw (8a) to (8);
	
	 \draw (4) to (5);
	 \draw (5) to (6);
	 \draw (6) to (7);
	 \draw (7) to[out=225, in=-45] (4);
	 \draw (7) to[out=135, in=45] (5);
	 \draw (4) to[out=45, in=135] (6);
    %(5)  [->] edge node [left] {} (2);
	
	\end{tikzpicture}
	\caption{$\underline{\delta}\in[\delta^\ast,\delta^{\ast\ast})$}
    \end{subfigure}
    \begin{tikzpicture}[
	contributor/.style={diamond,draw,thick, fill=blue!20, scale=0.7},
	periphery/.style={circle,draw,thick, scale=0.7, fill=yellow!20},
	core/.style={regular polygon,regular polygon sides=5,draw,thick, fill=green!20, scale=0.7},
	]
	%\matrix [draw,below left] at (current bounding box.north west) {
  %\node [contributor,label=right:Contributor] {}; &
  %\node [core,label=right:Core] {}; &
  %\node [periphery,label=right:Periphery] {}; \\
%};
    \end{tikzpicture}
\caption{$f(\cdot,\cdot,\cdot)=\frac{\lambda_i^\phi}{\lambda_i^\phi+\lambda_j^\phi}-\frac{1}{2}+\beta\lambda^\alpha_{ij}$, $c=1$, $\tau=1$}
\label{fig-propiib}
\end{figure}

Figures \ref{fig-propiia} and \ref{fig-propiib} sketch equilibria for different dispute technologies. Circles represent agents who share allies with their opponents, whereas other shapes and colors are clique specific. Increasing the benefits of cohesion transforms an overlapping society (Panel (a) of Figure \ref{fig-propiia}) into a segregated society (Panel (b) of Figure \ref{fig-propiia}) by encouraging agents to choose more mutual opponents with their allies. For high effectiveness of cohesion, we obtain a society consisting of two cliques as displayed in Panel (a) of Figure \ref{fig-propiib}. %In particular, the panel displays an equilibrium consisting of two cliques. 
Reducing the benefits from cohesion also decreases the incentives to choose similar allies as one's allies. The equilibrium network then consists of more than two cliques, which are also more homogeneous cliques (Panel (b) of Figure \ref{fig-propiib}).

Our simple equilibrium characterization uncovers important economic implications. Tolerance intervals are generally asymmetric in the BE. The value of a connection depends on the entire network in our model because cohesion depends on the allies of allies. %of the respective opponents are. 
Agents thus have an incentive to choose, at least to some extent, similar allies as their allies, thereby increasing their cohesion and their benefits from disputes with opponents. %Since we have made no particular assumption on the distribution of types (aside from continuity which serves to circumvent equilibrium multiplicity), this result is not an artifact of any specific realization of ideologies or how they are distributed. 
Proposition \ref{proposition:structural-balance} illustrates why there is no one-to-one relationship between agents' ideological proximity and their probability of being allies. %Instead, the allies of allies play an important role as well since they determine the cohesion an agent can achieve. 
In fact, agents with almost identical ideologies may be opponents and at the same time ally themselves with individuals whose ideologies are much farther from their own. 

This result rationalizes several seemingly paradoxical, yet common phenomena, for instance, (i) why extremist figures are able to secure support from moderates within their organization, (ii) why seemingly small ideological differences between organizations or factions prevent cooperation between them, or (iii), why seemingly small ideological differences within a political organization can cause its split. %, (iv) why seemingly niche movements have a substantial weight in the public debate and contribute a lot to polarization. 
To go beyond those insights, the next section investigates how the equilibrium structure and the associated outcomes are affected in a changing economy. %by changes to the parameters of the model.

\section{Dispute technology and polarization}\label{main}

The equilibrium characterization uncovers several important features about interactions among ideologically differentiated agents. A natural next step is to investigate which changes we might expect in the alliance network in a changing economy and the implication for polarization. To do so, we study the effect of an increase in the effectiveness of cohesion. %dispute technology alter the network structure and polarization in society. 
Cohesion may become more effective either directly, for instance because potential supporters value cohesion more, or indirectly, because the cost of effort determines how much benefits agents can extract from disputes through their strengths. Whenever it is costly to strengthen the ties to allies, agents become more homogeneous in their strengths and the relative benefits from cohesion increase. Indeed, many countries regulate campaigning expenditures or the contents that campaigns may contain to prevent coordination between political agents. Moreover, anti-lobbying efforts increase the hurdles for exchanges between politics and interest groups.%, thereby increasing the hurdles for cooperation between them.

Dispute intensity (our measure of polarization) depends on two characteristics of the network: (i) the number of disputes in society, i.e., the extensive margin of dispute intensity, and (ii) the strength of players in dispute captured by the sum of the weights of all their connections, i.e., the intensive margin of dispute intensity. It is thus useful to first establish a general result on the relationship between equilibrium efforts and equilibrium degrees.

\begin{lemma}\label{lemma:socialization-efforts}
    In the BE of the game, if $i$ has more neighbors than $j$ ($k^\ast_i\geq k^\ast_j$), then $i$ exerts lower effort than $j$ ($x^\ast_i\leq x^\ast_j$).
\end{lemma}

Lemma \ref{lemma:socialization-efforts} links a player's equilibrium degree to their equilibrium effort. Players with more allies exert lower efforts in equilibrium for two reasons. First, the network determines who are allies and opponents, so agents with more allies are necessarily in fewer disputes. Their efforts thus generate benefits in fewer instances, thereby reducing the incentives to exert effort. Second, efforts of neighbors also contribute to the strength of an agent. High degree players enjoy more spillovers from their neighbors' efforts and need not exert as much effort themselves. %in equilibrium. Such crowding out resembles similar strategic effects in the empirical literature, for instance in the context of fighting efforts of allied militias \citep{konig2017networks}. 

Lemma \ref{lemma:socialization-efforts} Lemma is useful to analyze the comparative statics of our model because it %establishes a link between the number of disputes (the extensive margin of dispute intensity) and the effort profile (the intensive margin of dispute intensity). The intensive margin of dispute intensity measures the strengths of players in dispute, which in turn depends on the effort profile. Lemma \ref{lemma:socialization-efforts} 
tells us when the extensive and intensive margin go in the same direction. 

%To build stronger intuition on the effects of interventions,
From now on, let us consider the parametric dispute technology of Definition \ref{definition:}, i.e., the augmented Tullock contest. We are interested two key comparative statics. First, how does a direct increase in the value of cohesion affect polarization? Second, how does an increase in the cost of exerting effort affect polarization? %We begin by summarizing the effect of a direct intervention on the dispute technology. 

\begin{proposition}\label{proposition:polarization}
	As the effectiveness of cohesion increases ($\beta$ or $\alpha$ increase),
	
	    \begin{enumerate}[{(}i{)}]
	        \item dispute intensity decreases if the network consists of cliques ($\delta^\ast>\underline{\delta}$);% and there exists $\Delta_1>0$ and $\Delta_0>0$, such that for all $i$, with $\theta_i\in[0,\Delta_0]$, or $\theta_i\in[1-\Delta_1,1]$, $x^\ast_i$ decreases; 
	        
	       \item otherwise, dispute intensity increases.
	    \end{enumerate}
\end{proposition}
 
Increasing the effectiveness of cohesion has ambiguous effects on polarization depending on whether the initial network is segregated or not. In an overlapping society, higher effectiveness of cohesion gives incentives for agents to coordinate on more similar neighborhoods with their allies. %Then, agents increase their cohesion since fewer of their allies are connected to some of their opponents.
There are consequently fewer overlaps in opponents' neighborhoods which leads to an increase in the number of disputes in the economy.\footnote{This is because tolerance is increasingly costly in the distance between the own and others' ideologies. Hence, agents choose narrower tolerance intervals in an overlapping society when cohesion becomes more effective.} An increase in polarization follows immediately because efforts are increasing in the number of disputes they are in (Lemma \ref{lemma:socialization-efforts}). Polarization increases unambiguously on the extensive and intensive margin and the first statement of Proposition \ref{proposition:polarization} follows. 

The reverse is true in an initially segregated society. As players derive greater benefits from cohesion, at least some cliques must become larger. The players in large cliques are in fewer disputes and exert lower efforts. This results in a decrease in dispute intensity on the extensive and intensive margin, thereby unambiguously decreasing polarization.

The results of Proposition \ref{proposition:polarization} illustrate the important role of the initial network structure for the effect of interventions in this economy. In general, there is no universally best tool to tackle polarization in ideologically heterogeneous societies. Instead, a similar intervention has drastically different effects in an overlapping society compared to a segregated society. %This highlights the necessity of carefully evaluating the initial conditions of the economy before recommending interventions to combat polarization.

An interesting insight of Proposition \ref{proposition:polarization} is that small groups of ideologically extreme types %, typically responsible for a substantial share of hostilities and therefore contributing substantially to polarization, 
may be a ``necessary evil" for a society. Attempts to dissolve such extremist group may in fact backfire and result in more polarization. This is because reducing the effectiveness of cohesion would at first reduce the incentive to be part of the ``silent majority", and encourage agents to join an ideologically extreme group. 

%Directly altering the effectiveness of cohesion, however, seems difficult and costly to implement. 
Various characteristics of a society, e.g., the media landscape, the general public, or even the characteristics of officials in parties or organizations, influence how effective cohesion is for political agents. A more tangible policy lever is to influence the cost of exerting effort. % in the ways discussed above. %Legislators may simply restrict the amount and type of communication between agents or groups of agents, for instance, restricting campaigning expenditures, regulating the content of campaigns, or anti-lobbying efforts tailored to increase the hurdles to exchange between parties and activist organizations. %Several countries also impose strict guidelines on campaigning, thereby increasing the hurdles to emphasize similarities between allied groups. 
%Through the eyes of our model, this corresponds to an increase in the cost of exerting effort, $c$, which is the focus of the next proposition.
To serve, exposition, we employ the notion of a \textit{dense society}. Formally, define a \textit{dense society} such that $n\to\infty$ and $P(\exists i\in N:\theta_i\in[\theta-\epsilon,\theta+\epsilon])\to1$.
%Proposition \ref{proposition:information cost} summarizes the effects of an increase in the cost of exerting effort. %has on dispute dispute intensity.%addresses the effects on dispute intensity and total socialization.
\begin{proposition}\label{proposition:information cost}
    If effort costs increase, total effort decreases. %Suppose there are benefits from cohesion (e.g., $\beta>0$). 
    In the parametric model with $\beta>0$ in a dense society, there exists a threshold $\tilde{\delta}<\delta^\ast$, such that
    \begin{enumerate}[{(}i{)}] 
        \item dispute intensity increases if the overlaps in players' neighborhoods are initially small, i.e., for an initially intermediate effort cost that induces $\delta^\ast>\underline{\delta}>\tilde{\delta}$;
        \item otherwise, dispute intensity decreases. %Moreover, there exist $\Delta_0>0$ and $\Delta_1>0$, such that for all $i$, with $\theta_i\in[0,\Delta_0]$, or $\theta_i\in[1-\Delta_1,1]$, $x^\ast_i$ decreases.
    \end{enumerate}
\end{proposition}

Increasing the cost of effort always reduces the total effort exerted by agents in equilibrium. While intuitive, this result is not entirely trivial. We must consider how potential changes in the network structure affect individual equilibrium efforts since changes in the effort cost also influence the indirect benefits from cohesion. The returns to exerting effort are diminishing, so even agents who engage in more disputes need not exert as much effort when their opponents' efforts tend to be small, for instance, due to an increase in $c$. Total efforts therefore decrease as exerting effort becomes costlier. This result establishes the effect of an increase in the effort cost on the intensive margin of dispute intensity, which is negative. The total effect on dispute intensity is \textit{ex ante} ambiguous since increasing the effort cost also alters agents' tolerance choices. 

Whenever an increase in the effort cost decreases the number of disputes, dispute intensity unambiguously decreases. Proposition \ref{proposition:information cost} identifies precisely when this is the case. For sufficiently high initial costs of exerting effort, agents are of relatively similar strength, even when their degrees differ. Poorly connected agents are then (almost) on par with better connected agents in terms of their strengths. %The benefits from an agent's strength diminish in equilibrium. 
While it becomes costlier for agents to establish reliable alliances, %which favours the poorly connected players who have few allies anyways. 
cohesive players can still extract high benefits. % from this game. %, since benefits from cohesion and and the cost of tolerance remain unaffected. 
In other words, agents substitute away from investing in their strengths towards investing in their cohesion. %An agent who invests in their cohesion chooses allies who are opponents of their opponents and segregated networks arise when effort is sufficiently costly. 
An increase in the effort cost would then imply that at least one clique becomes larger. %Greater benefits from cohesion push towards larger cliques when the initial network is segregated. 
The agents in larger cliques are in fewer disputes, thereby reducing the number of disputes in the economy. Dispute intensity decreases on the extensive margin %The effect of increasing the effort cost on the intensive margin of dispute intensity is negative. 
and the total effect is negative. %unambiguously decreases in $c$ for initially high effort costs. Vaguely speaking, high costs are such that the equilibrium network is segregated. 

%To understand the result for low costs of exerting effort, let us first consider the case of intermediate effort costs as this will also determine what we mean by ``low costs". 
Whenever the initial cost of exerting effort is not too high, the network is an overlapping society. Agents are sufficiently dissimilar in their strengths to ensure enough benefits stem from investing in their connections relative to investing in their cohesion. An increase in the cost of exerting effort reduces equilibrium efforts and thus makes agents more similar in terms of their strengths. The equilibrium network moves towards segregation. This increases the number of disputes. The effects on the extensive and intensive margin go in opposing directions. When the network is initially close to segregation, i.e., the cost of exerting effort is intermediate, increasing the cost of effort crowds out efforts only by little. This is because agents are in many disputes already and engage in even more due to the increase in the effort cost. Since agents in more disputes exert higher efforts (Lemma \ref{lemma:socialization-efforts}), the effect on the intensive margin becomes smaller as the network approaches segregation. The extensive margin eventually outweighs the intensive margin and dispute intensity increases in the cost of effort for an initially intermediate effort cost.%In other words, as agents substitute away from effort towards more cohesion, the crowding out of efforts becomes weaker for higher initial effort costs since the higher number of disputes creates a counter force. Note, 
\footnote{We assume a dense society in order to deliver a clean characterization of these effects. In a sparse society, similar results would occur, however, the agents would not alter their tolerance choices for marginal increases in the cost of exerting effort.} Assuming a dense society ensures the existence of an interval corresponding to ``intermediate effort cost". %must always exist in a dense society. The effect on the intensive margin of dispute intensity due to a marginal increase of the effort cost vanishes, while agents engage in more disputes, even for small increases in $c$.

Cases where the network is far from segregation correspond to ``low costs" of exerting effort. In this case, the extensive and intensive margin of dispute intensity also go in opposing directions when the cost of exerting effort increases. However, the intensive margin of dispute intensity outweighs the extensive margin. %, since the crowding out of efforts slows down only for higher effort costs. 
In some sense, the intermediate cost case emerges once the effect on the number of disputes dominates the crowding out of efforts. 

%This result also highlights an inherent difficulty with intervening in an ideologically diverse society. Suppose a planner wanted to infer the appropriate policy to dampen dispute intensity from the network structure. While the planner could easily distinguish between a high and an intermediate effort cost, it is not straightforward to infer whether the effort costs are low or intermediate from the network. However, o
Overlaps in the neighborhoods of allies are informative about the state of the world, i.e., whether the effort cost is low, high, or intermediate. This is important when trial and error strategies to elicit whether the economy is initially in the low cost or intermediate cost case are unfeasible. %Proposition \ref{proposition:information cost} provides some guidance on how to infer the appropriate policy from the network structure. In relatively segregated societies, reducing hurdles to effort reduces polarization.

%Figure \ref{infograph} illustrates an example.

\begin{figure}[ht]
	
    \centering
    \begin{subfigure}[b]{0.32\linewidth}
\centering
    \begin{tikzpicture}[
	contributor/.style={diamond,draw,thick, fill=blue!20, scale=0.7},
	periphery/.style={circle,draw,thick, scale=0.7, fill=yellow!20},
	core/.style={regular polygon,regular polygon sides=5,draw,thick, fill=green!20, scale=0.7},
	isolated/.style={regular polygon,regular polygon sides=6,draw,thick, fill=red!20, scale=0.7},
	]
	
	\path[draw,thick,below] (0,0) edge node {$\theta$} (3.5,0);
	\path[draw,thick,left] (0,0) edge node {$x_i$} (0,3.5);
	\node[periphery] at (3.5,2) (10){1};
	\node[periphery] at (0,2) (0){0};
	\node[periphery] at (3.2,1.5) (9){};
	\node[periphery] at (3,1.5) (8){};
	\node[periphery] at (2.7,1) (8a){i};
	\node[periphery] at (2.1,1.3) (7){};
	\node[periphery] at (1.8,1.3) (6){};
	\node[periphery] at (1.6,1.3) (5){};
	\node[periphery] at (0.3,1.5) (1){};
	\node[periphery] at (0.5,1.5) (2){};
	\node[periphery] at (0.9,1) (3){j};
	\node[periphery] at (1.3,1.3) (4){};

	\draw (0) to (1);
	\draw (1) to (2);
	\draw (2) to (3);
	\draw (3) to (4);
	\draw (4) to (5);
	\draw (5) to (6);
	\draw (6) to (7);
	\draw (7) to (8a);
	\draw (8a) to (8);
	\draw (8) to (9);
	\draw (9) to (10);
	 
	\draw (0) to[out=0, in=90] (2);
	\draw (3) to[out=180, in=-90] (1);
	 %\draw (3) to[out=225, in=-45] (0);
	
	\draw (9) to (10);
	\draw (8) to (9);
	\draw (8) to[out=90, in=180] (10);
	 %\draw (8a) to[out=-45, in=-135] (10);
	\draw (8a) to[out=0, in=-90] (9);
	\draw (8a) to (8);
	
	\draw (4) to (5);
	\draw (5) to (6);
	\draw (6) to (7);
	\draw (7) to[out=225, in=-45] (4);
	\draw (7) to[out=135, in=45] (5);
	\draw (4) to[out=45, in=135] (6);
	\end{tikzpicture}
	\caption{Intermediate $c$}
	\end{subfigure}
    \begin{subfigure}[b]{0.32\linewidth}
    \centering
    \begin{tikzpicture}[
	contributor/.style={diamond,draw,thick, fill=blue!20, scale=0.7},
	periphery/.style={circle,draw,thick, scale=0.7, fill=yellow!20},
	core/.style={regular polygon,regular polygon sides=5,draw,thick, fill=green!20, scale=0.7},
	isolated/.style={regular polygon,regular polygon sides=6,draw,thick, fill=red!20, scale=0.7}
	]
	\path[draw,thick,below] (0,0) edge node {$\theta$} (3.5,0);
	\path[draw,thick,left] (0,0) edge node {$x_i$} (0,3.5);
	\node[contributor] at (3.5,1.2) (10){1};
	\node[isolated] at (0,1.2) (0){0};
	\node[contributor] at (3.2,1.2) (9){};
	\node[contributor] at (3,1.2) (8){};
	\node[contributor] at (2.7,1.2) (8a){};
	\node[core] at (2.1,1.2) (7){};
	\node[core] at (1.8,1.2) (6){};
	\node[core] at (1.6,1.2) (5){};
	\node[isolated] at (0.3,1.2) (1){};
	\node[isolated] at (0.5,1.2) (2){};
	\node[isolated] at (0.9,1.2) (3){};
	\node[core] at (1.3,1.2) (4){};
	
	 %\draw (0) to (1);
	 %\draw (1) to (2);
	 \draw (2) to (3);
	 \draw (0) to[out=45, in=135] (2);
	 \draw (3) to[out=135, in=45] (1);
	 \draw (3) to[out=225, in=-45] (0);
	
	 \draw (9) to (10);
	 \draw (8) to (9);
	 \draw (8) to[out=45, in=135] (10);
	 \draw (8a) to[out=-45, in=-135] (10);
	 \draw (8a) to[out=45, in=135] (9);
	 \draw (8a) to (8);
	
	 \draw (4) to (5);
	 %\draw (5) to (6);
	 \draw (6) to (7);
	 \draw (7) to[out=225, in=-45] (4);
	 \draw (7) to[out=135, in=45] (5);
	 \draw (4) to[out=45, in=135] (6);
	\end{tikzpicture}
	\caption{$\underline{\delta}\in [\delta^\ast,\delta^{\ast\ast})$}
	\end{subfigure}
    \begin{subfigure}[b]{0.32\linewidth}
    \centering
    \begin{tikzpicture}[
	contributor/.style={diamond,draw,thick, fill=blue!20, scale=0.7},
	periphery/.style={circle,draw,thick, scale=0.7, fill=yellow!20},
	core/.style={regular polygon,regular polygon sides=5,draw,thick, fill=green!20, scale=0.7},
	isolated/.style={regular polygon,regular polygon sides=6,draw,thick, fill=red!20, scale=0.7},
	]
	\path[draw,thick,below] (0,0) edge node {$\theta$} (3.5,0);
	\path[draw,thick,left] (0,0) edge node {$x_i$} (0,3.5);
	\node[contributor] at (3.5,0) (10){1};
	\node[isolated] at (0,0) (0){0};
	\node[isolated] at (3.2,0) (9){};
	\node[isolated] at (3,0) (8){};
	\node[isolated] at (2.7,0) (8a){};
	\node[isolated] at (2.1,0) (7){};
	\node[isolated] at (1.8,0) (6){};
	\node[isolated] at (1.6,0) (5){};
	\node[isolated] at (0.3,0) (1){};
	\node[isolated] at (0.5,0) (2){};
	\node[isolated] at (0.9,0) (3){};
	\node[isolated] at (1.3,0) (4){};
	
	 %\uncover<3->{\draw (0) to (1);}
	 %\uncover<3->{\draw (1) to (2);}
	 %\uncover<3->{\draw (2) to (3);}
	 %\uncover<3->{\draw (0) to[out=45, in=135] (2);}
	 %\uncover<3->{\draw (3) to[out=135, in=45] (1);}
	 %\uncover<3->{\draw (3) to[out=225, in=-45] (0);}
	
	%\uncover<3->{\draw (0) to (1);}
	%\uncover<3->{\draw (1) to (2);}
	%\uncover<3->{\draw (2) to (3);}
	%\uncover<3->{\draw (3) to (4);}
	%\uncover<3->{\draw (4) to (5);}
	%\uncover<3->{\draw (5) to (6);}
	%\uncover<3->{\draw (6) to (7);}
	%\uncover<3->{\draw (7) to (8a);}
	%\uncover<3->{\draw (8a) to (8);}
	%\uncover<3->{\draw (8) to (9);}
	\draw (0) to[out=45, in=135] (9);
	\end{tikzpicture}  
	\caption{$\underline{\delta}\geq\delta^{\ast\ast}$}  
	\end{subfigure}
    \caption{$f(\cdot,\cdot,\cdot)=\frac{\lambda_i^\phi}{\lambda_i^\phi+\lambda_j^\phi}-\frac{1}{2}+\beta\lambda^\alpha_{ij}$, $c=1$, $\tau=1$}
	\label{infograph}
\end{figure}
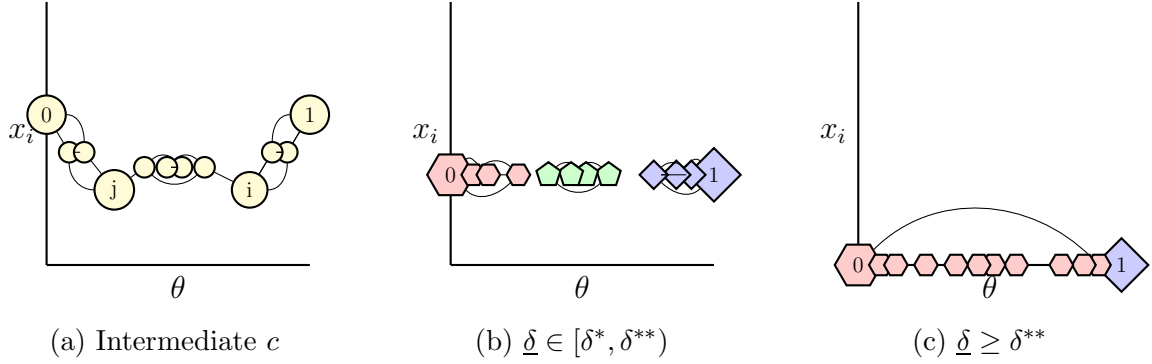

Figure \ref{infograph} sketches the transformations of the BE network when effort costs increase from intermediate to high. %An overlapping society turns into a segregated society as $c$ increases. 
Nodes $i$ and $j$ connect otherwise segregated players in an overlapping society (Panel (a)). An increase in the effort cost increases the relative benefits from cohesion. Player $i$ ($j$) thus benefits more from being in dispute with everyone to their left (right). This is because all players to their right (left) are in dispute with the same players. %A segregated network arises. 
The cohesion of players $i$ and $j$ is higher in Panel (b) compared to Panel (a) and so is the cost of tolerance. A further increase in the effort cost increases the size of the largest clique and we ultimately arrive in the situation displayed in Panel (c). %\ref{proposition:information cost}. 

\section{The appeal of extremists}\label{extremism}

Heterogeneity in ideologies always raises the question of the role of ``extreme ideologies". In our model, extreme ideologies are types close to zero or one. %Fundamentally, ideology introduces homophily to our model. Homophily influences the \textit{appeal} of extremists through two channels. First,
On the one hand, homophily makes extreme ideologies unappealing, since allying oneself with extreme ideologies induces a high reputation loss for most types. On the other hand, %dispute technology becomes more effective in the cohesion of agents. In this sense, homophily increases the appeal of extreme ideologies since 
agents with extreme ideologies are in many disputes, thereby substantially increasing the cohesion of their allies. Extreme types are thus appealing to those who are sufficiently close to them. How these agents respond to changes to the relative benefits from cohesion is crucial to understand the overall effects of attempting to influence extreme types. %\footnote{We are waving our hands here on what ``extreme" ideologies actually are in this context. Ultimately, this depends on the distribution of ideologies. Nonetheless, even in a society where far left and far right ideologies are prevalent, there are always those who are even more extreme.} %On the other hand, agents of extreme ideologies are in many disputes since they are more restricted in what direction to tolerate. Such agents are therefore in many disputes and allying oneself with extremists allows players to differentiate from their opponents. They consequently dispute more effectively against their opponents. 
The next corollary addresses this. %presents a result on the role of extreme ideologies. 

\begin{corollary}\label{cor:heterogeneity}
    In a dense overlapping society ($\delta^\ast>\underline{\delta}$), if the benefits from cohesion ($\underline{\delta}$) increase, extreme types exert lower effort and are in fewer disputes.
\end{corollary}

The behavior of players with extreme ideologies is generally ambiguous when cohesion becomes more effective in segregated societies (Proposition \ref{proposition:polarization} and \ref{proposition:information cost}). In principle, as one clique grows, extreme types may lose connections and increase their efforts to dispute against others. In an overlapping society, however, those extreme types become increasingly appealing as connections since they are in many disputes and cohesion becomes more effective. Then, players with extreme allies dispute effectively against their opponents and extreme types form more connections. Those types also reduce their efforts and engage in fewer disputes, while overall polarization increases. 

The results of Corollary \ref{cor:heterogeneity} rationalize why relatively moderate types may be drawn towards the ideologically extreme and away from ideologically more similar moderates. Extreme allies serve to show a clear edge towards others and do so more effectively compared to moderate allies. %This creates a temptation to interact with extremists.  

Corollary \ref{cor:heterogeneity} also uncovers a trade-off between ``more polarization" and ``influencing extremists". Against common intuition, interventions tailored to reduce the influence of few ``extremists" on polarization may in fact end up increasing polarization in society. This is because such extremists continue to polarize by dragging more moderate types into their tolerance window, who in turn contribute more to overall polarization.\footnote{This result exhibits similarities to the case of intolerant extremists in \cite{genicot2022tolerance}. When extremists are less tolerant, moderate types compromise for extremists and consequently choose extreme actions.} %Polarization arises. In our model and under certain circumstances, moderates tolerate extremists to increase their cohesion.

\section{Discussion}\label{discussion}

In this section, we discuss some extensions and implications of the model. The appropriate proofs and other extensions can be found in Appendix \ref{appendixB}. %This includes implications of the baseline analysis as well as meaningful extensions that preserve the results of the paper. First, we present some additional results of the baseline model.

\textbf{Initiation of dispute:} In the baseline model, tolerance decisions also directly determine disputes. While natural for the contexts we have in mind, this assumption does not qualitatively affect the results of the paper. Suppose agents pay a finite cost for initiating a dispute with someone, denoted by $D$. Let $g_{ij}=-1$ denote a dispute between $i$ and $j$, whereas $g_{ij}=0$ indicates a neutral relationship. It is natural to think of dispute initiation as a one-sided decision, so let $\bar{g}_{ij}=-1$ if $\min\{g_{ij},g_{ji}\}=-1$.%A neutral relationship between $i$ and $j$ is then denoted by $g_{ij}=0$. A network $\bar{g}$ describes connections, disputes and neutral relationships among players.
%\footnote{Note, (positive) connections are by construction symmetric in our model and we can simply take $g_{ij}=\bar{g}_{ij}$ whenever $\bar{g}_{ij}\neq-1$.} %Let $\lambda_i(\bar{g})=\sum_{h\in K_i(\bar{g})}g_{ih}$ be the strength of player $i$, i.e., the sum of positive link weights. 

\begin{proposition}\label{proposition:prize}
    If cohesion is beneficial in a dense society, there exists a threshold $\underline{D}$, such that for a dispute initiation cost $D\leq\underline{D}$, each agent $i$ is in dispute with everyone outside their interval of tolerable types, $\theta_j\not\in[\underline{t}_i,\bar{t}_i]$.
\end{proposition}

A dispute is initiated if the benefits outweigh the costs. The game is economically equivalent to the baseline model.\footnote{However, some technical inconveniences arise. First, agents would want to fend off (some) disputes by investing a lot into their strength, i.e., choosing high efforts. However, given disputes against them are not initiated, agents would want to reduce their efforts \textit{ex post}. Non-existence of an equilibrium may arise.} Sufficiently low costs ensure the initiation of each dispute.\\ %is initiated and no neutral relationships remain. However, choosing a weaker solution concept would possibly also ensure equilibrium existence. 

%Moreover, dispute initiation can create equilibrium multiplicity when two agents want to initiate a dispute against each other. Since it is costly to initiate a dispute, $D>0$, both agents would prefer the other agent to initiate the dispute to save the cost of initiation. However, agents may prefer to initiate the dispute if the other player does not. Appropriate alternative assumptions would then potentially create similar results to ours.

\textbf{Heterogeneous flexibility:} The baseline model assumes uniform flexibility $\tau$. Allowing for any possible $\tau_i\ne\tau_j$ is difficult to characterize due to its many degrees of freedom. Suppose instead flexibility correlates with ``extremism". We study two versions of the model with either \textit{stubborn} or \textit{flexible} extremists.\footnote{Extremists hold strong views and consequently insist on them, making them less willing to tolerate other ideologies. Indeed, \cite{van2017extreme} document \textit{dogmatic intolerance} of individuals with extreme political beliefs. On the other hand, extremists may be more grounded in their ideology and therefore appreciate interactions with those of different ideologies.}\footnote{\cite{genicot2022tolerance} uses a similar approach to impose structure on agents' willingness to compromise for others.} 

To introduce stubborn extremists formally, let $\tau_i=\underline{\tau}+\kappa\vert\theta_i-1/2\vert$, i.e., more extreme players pay a higher cost for tolerance. 

The model with stubborn extremists preserves orderedness of the BE network. Moderates are less constraint in their tolerance decision, since they may tolerate ideologies to their right and their left. Stubborn extremists would therefore never tolerate other extremists, but only more flexible moderates. The flexibility of agents with extreme ideology is thus decisive for whether they connect to moderates and the equilibrium network is ordered. The model with stubborn extremists thus produces qualitatively similar results to the baseline model, however, we would expect segregation to arise for lower benefits from cohesion.  

Whereas stubborn extremists seem to be the natural assumption, sometimes extremists are more flexible in whom they want to interact with. Consider the specification $\tau_i=\max\{\bar{\tau}-\kappa\vert \theta_i-1/2\vert,0\}$, which is the model with \textit{flexible} extremists. 

Since the stubborn moderates tend to be intolerant towards flexible extremists, there may arise situations where extremists tolerate extremists from the other end of the ideological spectrum. The BE network with flexible extremists need no longer be ordered and other equilibrium structures may exist. The characterization of these structures is beyond the scope of this paper. However, we could imagine, for appropriate values of $\kappa$, $\alpha$ and $\beta$, a segregated equilibrium where agents with extreme ideology tolerate those of opposing end of the ideology space while no one else tolerates them. Stubborn moderates leave extreme types with too few allies who then resort to becoming tolerant towards extremists instead. %Figure \ref{fig:tolerant-extremists} illustrates an example. 

\begin{figure}[ht]
    \centering
    \begin{tikzpicture}[
	contributor/.style={diamond,draw,thick, fill=blue!20, scale=0.7},
	periphery/.style={circle,draw,thick, scale=0.7, fill=yellow!20},
	core/.style={regular polygon,regular polygon sides=5,draw,thick, fill=green!20, scale=0.7},
	isolated/.style={regular polygon,regular polygon sides=6,draw,thick, fill=red!20, scale=0.7},
	]
	\path[draw,thick,below] (0,0) edge node {$\theta$} (14,0);
	\path[draw,thick,left] (0,0) edge node {$x^\ast_{i}$} (0,5);
	\node[contributor] at (14,4) (10){1};
	\node[contributor] at (0,4) (0){0};
	\node[isolated] at (12.5,1) (9){};
	\node[isolated] at (11.6,1) (8){};
	\node[isolated] at (10.4,1) (8a){};
	\node[isolated] at (8.4,1) (7){};
	\node[isolated] at (7.2,1) (6){};
	\node[isolated] at (6,1) (5){};
	\node[isolated] at (1,1) (1){};
	\node[isolated] at (2,1) (2){};
	\node[isolated] at (3,1) (3){};
	\node[isolated] at (5,1) (4){};
	
	\draw (0) to (10);
	\draw (1) to (2);
	\draw (2) to (3);
	\draw (3) to (4);
	\draw (4) to (5);
	\draw (5) to (6);
	\draw (6) to (7);
	\draw (7) to (8a);
	\draw (8a) to (8);
	\draw (8) to (9);
	%\draw (0) to[out=45, in=135] (9);
	
	\end{tikzpicture}
    \begin{tikzpicture}[
	contributor/.style={diamond,draw,thick, fill=blue!20, scale=0.7},
	periphery/.style={circle,draw,thick, scale=0.7, fill=yellow!20},
	core/.style={regular polygon,regular polygon sides=5,draw,thick, fill=green!20, scale=0.7},
	]
	%\matrix [draw,below left] at (current bounding box.north west) {
  %\node [contributor,label=right:Contributor] {}; &
  %\node [core,label=right:Core] {}; &
  %\node [periphery,label=right:Periphery] {}; \\
%};
    \end{tikzpicture}
\caption{$f(\cdot,\cdot,\cdot)=\frac{\lambda_i^\phi}{\lambda_i^\phi+\lambda_j^\phi}-\frac{1}{2}+\beta\lambda^\alpha_{ij}$, $c=1$, $\tau=2\kappa$. In the BE network, $1$ and $0$ are in dispute with everyone but each other and tolerate players that they are in dispute with.}
\label{fig:tolerant-extremists}
\end{figure}
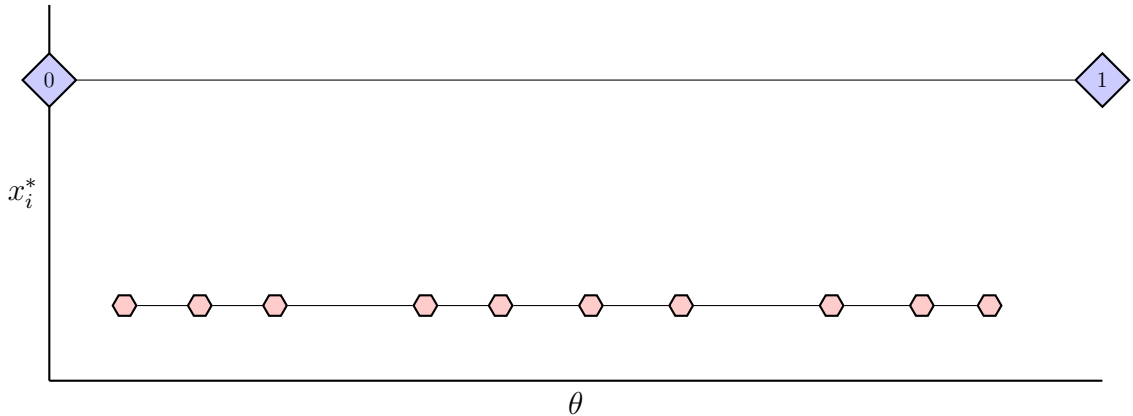

The network in Figure \ref{fig:tolerant-extremists} sketches an equilibrium consisting of two cliques in dispute. One clique contains the flexible extremists $1$ and $0$ whereas the other clique contains the remaining stubborn moderates. The tolerance cost for $1$ and $0$ is low enough to tolerate each other whereas moderates benefit enough from cohesion in order not to tolerate either extremist. %Obviously, the equilibrium network that this example displays is not ordered with respect to types.  

\section{Conclusion}\label{conclusion}

This paper studies how the concern for cohesion impacts the arising ally and opponent relationships among ideologically differentiated political actors. Agents choose the ideological composition of their allies by their tolerance decision and the strength of their connections by a generic effort. %Exerting tolerance and effort is costly, however, increases an agent's strength and cohesion in dispute with their opponents, i.e., players who are not allies. 
The number of allies and the intensity of connections to them captures the strength of an agent and uniformly increases their benefits from a dispute with everyone who is not their ally. Moreover, an agent is more cohesive in a dispute and derives greater benefits from it when they share fewer allies with their opponent. 

The equilibrium network structure depends on how effective cohesion is. If cohesion is sufficiently effective, the equilibrium network is segregated into cliques. %Generally, for greater benefits from cohesion, larger cliques arise. 
Otherwise, the network is an overlapping society where opponents have mutual allies. Agents are generally not allies with those who are ideologically closest. Instead, agents of relatively similar ideologies may be opponents and have allies who are ideologically farther. 

An increase in the effectiveness of cohesion has ambiguous effects on the division in society. In an initially segregated network, at least some cliques become larger and thus ideologically diverse, thereby reducing polarization in society. For an initially overlapping society, on the other hand, higher concerns for cohesion lead to more segregation and polarization. 

Altering the cost of exerting effort is a tool for a policy maker to influence the relative benefits from cohesion. For higher effort costs, agents are more concerned with their cohesion and high enough effort costs induce a segregated equilibrium network. %The equilibrium network is segregated. 
Then, higher relative concerns for cohesion would lead more to diverse cliques and reduce polarization. 

For low and intermediate initial effort costs, an increase in the effort cost encourages segregation. Only for an intermediate effort cost, the network is initially close to segregation and the increase in the number of disputes outweighs polarizes more than the reduction of efforts decreases polarization.

Our results illustrate the complexity behind the incentives of choosing allies and opponents when pursuing an ideological agenda. In general, ideological proximity does not imply cooperation between agents. This sheds light on why some ideological organizations rally behind extremist candidates within or why seemingly close ideological agents and organizations are sometimes bitter rivals, even when they originate from the same ideas and convictions. 

\bibliography{library}

\newpage

\setcounter{theorem}{0}
\setcounter{proposition}{0}
\setcounter{corollary}{0}

\renewcommand{\theequation}{A-\arabic{equation}}
\renewcommand\theproposition{A-\arabic{proposition}}
\renewcommand\thecorollary{A-\arabic{corollary}}
\renewcommand\thetheorem{A-\arabic{theorem}}

\addcontentsline{toc}{section}{Appendices}
\renewcommand{\thesubsection}{\Alph{subsection}}

\appendixtitleon
\appendixtitletocon
\begin{appendices}

\section{Proofs}\label{appendix-A}

\begin{proof}[Proof of Lemma \ref{lemma:ordered}]
Suppose \textit{ad absurdum} $\theta_i>\theta_j$ and $\bar{t}_i<\bar{t}_j$ wlog. We distinguish two cases.

\noindent\textbf{Case 1:} $\underline{t}_i>\underline{t}_j$. For the highest type in $[\underline{t}_j,\bar{t}_j]$, say $h$, $\theta_h=\bar{t}_j$. Hence, $\theta_j\in[\underline{t}_h,\bar{t}_h]$, since $g_{ij}=0$ otherwise and narrowing $[\underline{t}_j,\bar{t}_j]$ is a profitable deviation. Moreover, $h>i$ and $\theta_h\not\in[\underline{t}_i,\bar{t}_i]$ whereas $\theta_h\in[\underline{t}_j,\bar{t}_j]$. Since $h>j$ by assumption, $\underline{t}_h\leq\theta_j$. Otherwise, $g^\ast_{jh}=0$, which would imply $j$ could increase their profit by decreasing $\bar{t}_j$. Moreover, $\bar{t}_i<\theta_h$ by assumption, which implies $\bar{t}_i=\bar{t}_j$ is a profitable deviation for $i$ when $\bar{t}_j\geq \theta_h$. Otherwise, $\bar{t}_j$ is not optimal. %If $\bar{t}_h<\theta_i$, it must be profitable for $h$ to increase $\bar{t}_h$ to $\bar{t}_h=\theta_h$, since $h>i$ implies a cost saving of $\tau(\theta_h-\bar{t}_h)^2$ without any loss in benefits. 
This contradiction addresses \textbf{Case 1}. 

\noindent\textbf{Case 2:} $\underline{t}_i<\underline{t}_j$. In this case, there exists $h>i$ and $\ell<j$ such that $\theta_\ell\in [\underline{t}_i,\bar{t}_i]$ and $\theta_h\in [\underline{t}_j,\bar{t}_j]$ while $\theta_h\not\in [\underline{t}_i,\bar{t}_i]$ and $\theta_\ell\not\in [\underline{t}_j,\bar{t}_j]$. Since tolerance is costly and neighbors generate indirect benefits from disputes, either $i$ can increase their payoff by choosing $\underline{t}_i=\underline{t}_j$ and $\bar{t}_i=\bar{t}_j$, or $j$ can increase their payoff by choosing $\underline{t}_j=\underline{t}_i$ and $\bar{t}_j=\bar{t}_i$, a contradiction to the initial assumption. %This is wlog, since a similar argument holds for $h$ and $\ell$ and we need not consider incentives of $h$ and $\ell$. 
This addresses \textbf{Case 2} the statement follows.
\end{proof}\\

\begin{proof}[Proof of Proposition \ref{proposition:existence}]
To establish existence, it is sufficient to construct an equilibrium. Take agent $1$ and fix $[\underline{t}_1,1]$. In equilibrium, $i\not\in[\underline{t}_1,1]$ implies $\bar{t}_i<1$, since $\tau>0$ and $g_{i1}=0$ independently of $[\underline{t}_i,\bar{t}_i]$ (Lemma \ref{lemma:ordered}). Similarly, let $0$ choose $[0,\bar{t}_0]$. Next, take $i>j$ for all $j\in N$ with $\theta_j<\underline{t}_1$ and let them choose $[\underline{t}_i,\bar{t}_i]$. By construction, no profitable deviation emerges for $1$, since $g_{i1}=0$ in any case. For each $i'>i$, $\underline{t}_{i'}>\underline{t}_i$ (Lemma \ref{lemma:ordered}). Let each $i'$ choose $[\underline{t}_{i'},\bar{t}_{i'}]$. By construction, no profitable deviation emerges. If $\underline{t}_i=0$, let each $j\in N$ choose $[\underline{t}_j,\bar{t}_j]$ and $x_j$ accordingly. Otherwise, take $h\leq j$ for all $j\in N$ with $\theta_j>\bar{t}_0$ and let them choose $[\underline{t}_h,\bar{t}_h]$. Continue this process until each player has chosen $[\underline{t}_i,\bar{t}_i]$. From our assumptions on $f(\cdot,\cdot,\cdot)$, for any given network, there is a unique profile of optimal efforts, $x$. Hence, letting each agent $i$ choose $x_i$ constitutes an equilibrium $s^\ast$. %Otherwise, take $j<\underline{t}_i$, where $j\geq h$ for all $h$, with $\theta_h<\underline{t}_i$ and repeat the process until, for some $m$, $\underline{t}_m=0$. Let each $i\in N$ choose $x_i$ accordingly. The resulting network constitutes an equilibrium, which proves the first statement of the proposition.
This establishes existence. 

\noindent Next, we address uniqueness of the bilateral equilibrium. First, note that for any configuration of $\mathbf{t}=\mathbf{t}_1\times...\times \mathbf{t}_n$, there exists a unique optimal effort vector $x$. %This is because $f(\cdot,\cdot,\cdot)$ is continuous, strictly concave, and increasing in its first argument as well as continuous, decreasing and concave in its second argument. 
Thus, the equilibrium is unique if there exists a unique $\mathbf{t}^\ast$. Suppose \textit{ad absurdum} there exist $s^\ast$ and $s^{\ast\ast}$ which constitute an equilibrium of the game. %By Proposition \ref{proposition:structural-balance}, any equilibrium network is ordered. 
By Lemma \ref{lemma:ordered}, there exists at least one player, say $i$, who is indifferent between tolerating $j$ and not tolerating $j$, or who is indifferent between tolerating $j$ or $h$. We address the cases separately and assume $h>i>j$ wlog. 

\noindent\textbf{Case 1}: Suppose $u_i(s^\ast)=u_i(s^{\ast\ast})$ where $\underline{t}^\ast_i=\theta_j$ and $\underline{t}^{\ast\ast}_i>\theta_j$, i.e., $i$ is indifferent between tolerating $j$ and not tolerating $j$. Denote by $\ell$ the lowest type whom $i$ tolerates in $s^{\ast\ast}$, i.e., $\underline{t}^{\ast\ast}_i=\theta_\ell$. %Denote by $g^\ast$ the equilibrium network where $j\in K_i(g^\ast)$ and $g^{\ast'}$ the equilibrium where $j\not\in K_i(g^{\ast'})$. 
Agent $i$ is indifferent if $\sum_{l\not\in K_i(g^\ast)}f(\lambda^\ast_i,\lambda^\ast_l,\lambda^\ast_{il})-\sum_{l\not\in K_i(g^{\ast\ast})}f(\lambda^{\ast\ast}_i,\lambda^{\ast\ast}_l,\lambda^{\ast\ast}_{il})=\tau[(\theta_j-\theta_i)^2-(\theta_\ell-\theta_i)^2]$, where $\theta_\ell>\theta_j$ by construction. By the assumptions on the distribution of types, this holds for some $i,j\in N$ with probability zero. Hence, no player $i$ is indifferent between tolerating and not tolerating another agent $j$.

\noindent\textbf{Case 2}: Suppose $u_i(s^\ast)=u_i(s^{\ast\ast})$ where $\underline{t}^\ast_i=\theta_j$ and $\bar{t}^{\ast}_i<\theta_h$ whereas $\underline{t}^{\ast\ast}_i>\theta_j$ and $\bar{t}^{\ast\ast}_i=\theta_h$, i.e., $i$ is indifferent between tolerating $j$ and tolerating $h$. By assumption, $P(\exists i,j,h\in N:\{\vert\theta_i-\theta_j\vert=\vert\theta_i-\theta_h\vert\})=0$. Either $j$ or $h$, say $h$, must have weakly more mutual opponents with $i$. This implies $\sum_{l\in K^\ast_i}f(\lambda^\ast_i,\lambda^\ast_l,\lambda^\ast_{il})-\sum_{l\in K^{\ast'}_i}f(\lambda^{\ast\ast}_i,\lambda^{\ast\ast}_l,\lambda^{\ast\ast}_{il})=\tau[(\theta_i-\theta_h)^2-(\theta_i-\theta_j)^2]$, i.e., the additional cost of tolerating $h$ instead of $j$ equals the additional benefits of tolerating $h$ instead of $j$. Since $f(\cdot,\cdot,\cdot)$ is discretely increasing in its third argument and by the assumptions on the distribution of types, this condition holds with probability zero, a contradiction that establishes uniqueness. This concludes the proof of Proposition \ref{proposition:existence}.
\end{proof}\\

\begin{proof}[Proof of Proposition \ref{proposition:structural-balance}]
%We prove the proposition by establishing a separate lemma first. This lemma establishes structure on tolerance decisions in the unique bilateral equilibrium.
By Lemma \ref{lemma:ordered}, it is sufficient to show that one can always find values of $\underline{\delta}$ such that players form cliques. Suppose \textit{ad absurdum} no $\delta^\ast$ and $\delta^{\ast\ast}$ exist. Take some $\underline{\delta}=\delta(n-2)-\delta(n-3)$ wlog. For the largest possible clique $C(g^\ast)$, $\vert C(g^\ast)\vert=n-1$. Suppose, $\underline{t}_1=\theta_i$ and $\bar{t}_i=1$, where $i\leq j$ for all $j\in N\setminus\{0\}$. By Lemma \ref{lemma:ordered}, it is sufficient to establish a contradiction for $1$. The cost saving for player $1$ from deleting a link to $i$ from the clique is $\tau[(1-\theta_j)^2-(1-\theta_i)^2]$, where $\nexists h$, such that $i<h<j$. This expression is finite, since $[0,1]$ is compact and bounded. %Moreover, $x_1^\ast$ is weakly increasing due to this deviation, since $f(\cdot,\cdot,\cdot)$ is concave and increasing in its first argument. 
This contradicts non-existence of $\delta^{\ast\ast}$. A network consisting of fewer cliques requires more tolerance and $\delta^{\ast\ast}\geq\delta^\ast$ follows immediately. %This implies existence of $\delta^\ast$. 
The characterization follows trivially, which concludes the proof of Proposition \ref{proposition:structural-balance}.
\end{proof}\\

\begin{proof}[Proof of Lemma \ref{lemma:socialization-efforts}]
Suppose \textit{ad absurdum} $k_i(g^\ast)>k_j(g^\ast)$ and $x^\ast_i>x^\ast_j$. We distinguish two cases.
    
\noindent\textbf{Case 1:} Suppose $\underline{\delta}\geq\delta^\ast$. Hence, the equilibrium network is segregated. If $j\in K_i(g^\ast)$ and $k_i=m$, then $k_j=m$. We must therefore consider the case where $g^\ast_{ij}=0$ in $s^\ast$. Suppose $k_j=m-1$ wlog. By construction, $j$ is in dispute with more players than $i$ since $k_j(g^\ast)<k_i(g^\ast)$. Hence, $\sum_{j\not\in K_i(g^\ast)}f'(\lambda_i,\lambda_\ell,\lambda_{i\ell})=\sum_{h\not\in K_j(g^\ast)}f'(\lambda_j,\lambda_h,\lambda_{jh})=c$. This is a contradiction, since $f(\cdot,\cdot,\cdot)$ is strictly concave in its first argument.

\noindent\textbf{Case 2:} Suppose $\delta^\ast>\underline{\delta}$. %Call $h$ the representative neighbor of $i$ and $\ell$ the representative neighbor of $j$. For $i$ with $k_i=m$ and $j$ with $k_j\leq m-1$, 
The first-order conditions require $\sum_{p\not\in K_i(g^\ast)}f'(\lambda_i,\lambda_p,\lambda_{ip})=c$ for all $i\in N$. For $k_i=m$ and $k_j\leq m-1$, there must exist $\ell\in K_j(g^\ast)$ and $\ell\not\in K_i(g^\ast)$ as well as $h\in K_i(g^\ast)$ and $h\not\in K_j(g^\ast)$ such that $x^\ast_h<x^\ast_\ell$. Otherwise, either $\sum_{p\not\in K_i(g^\ast)}f'(\lambda_i,\lambda_p,\lambda_{ip})<c$ or $\sum_{q\not\in K_j(g^\ast)}f'(\lambda_j,\lambda_q,\lambda_{jq})>c$, a contradiction. Moreover, for each such player $\ell$ and $h$, $\sum_{p'\not\in K_\ell(g^\ast)}f'(\lambda_\ell,\lambda_{p'},\lambda_{\ell p'})=\sum_{q'\not\in K_h(g^\ast)}f'(\lambda_h,\lambda_{q'},\lambda_{h q'})=c$. Hence, there must exist $h'$ such that $h'\in K_h(g^\ast)$ and $h'\not\in K_\ell(g^\ast)$, as well as $\ell'$ such that $\ell'\in K_\ell(g^\ast)$ and $\ell'\not\in K_h(g^\ast)$, and $x^\ast_{\ell'}<x^\ast_{h'}$, and so on.  

\noindent By Lemma \ref{lemma:ordered}, the BE network is ordered. Assume that $h'=1\geq h$ and $\ell' = 0$. Since Lemma \ref{lemma:ordered} tells us that the BE is ordered, any $h''\in (i,h)$ ($\ell''\in (\ell,j)$) is a neighbor of $i$ and $h$ ($j$ and $\ell$) for whom $\sum_{p\not\in K_{h''}(g^\ast)}f'(\lambda^\ast_{h''},\lambda^\ast_{p},\lambda_{h''p})=c$ ($\sum_{q\not\in K_{\ell''}(g^\ast)}f'(\lambda^\ast_{\ell''},\lambda^\ast_{q},\lambda_{\ell''q})=c$). Moreover, we know that $K_0(g^\ast)\subseteq K_{\ell}(g^\ast)$ and $K_1(g^\ast)\subseteq K_{h}(g^\ast)$ and thus $x^\ast_1>x^\ast_v>x^\ast_h$ and $x^\ast_0>x^\ast_o>x^\ast_\ell$ for all $o\in(0,\ell)$ and $v\in(h,1)$ (Lemma \ref{lemma:ordered}). Hence, for each $h''\in(h,i)$, $x^\ast_h<x^\ast_{h''}<x^\ast_i$ and for each $\ell''\in(h,i)$, $x^\ast_\ell>x^\ast_{\ell''}>x^\ast_j$. Since $K_0(g^\ast)\subseteq K_\ell(g^\ast)$, i.e., $0$ is in more disputes than $\ell$, $x^\ast_\ell>x^\ast_j$ would imply that $k_\ell>k_j$. A similar argument for $1$, $h$ and $i$ implies $k_j<k_i<k_h$, where, $\sum_{q\not\in K_{i}(g^\ast)}f'(\lambda^\ast_{i},\lambda^\ast_{q},\lambda_{iq})=c=\sum_{p\not\in K_{h}(g^\ast)}f'(\lambda^\ast_{h},\lambda^\ast_{p},\lambda_{hp})$ and $\sum_{q\not\in K_{1}(g^\ast)}f'(\lambda^\ast_{1},\lambda^\ast_{q},\lambda_{1q})=c=\sum_{p\not\in K_{h}(g^\ast)}f'(\lambda^\ast_{h},\lambda^\ast_{p},\lambda_{hp})$. By Lemma \ref{lemma:ordered}, were $i$ and $h$ to delete their link, $\lambda^\ast_i$ decreases by less than $\lambda^\ast_h$, since $x^\ast_i>x^\ast_h$. Moreover, after deleting the connection, $k_h>k_i$ must still hold since $x^\ast_i>x^\ast_{h''}>x^\ast_h$ allows us to take $\bar{t}_i=\theta_h$ wlog. This implies that the initial condition $x^\ast_h<x^\ast_i$ could not have been part of a BE strategy%since $x^\ast_i>x^\ast_{h''}>x^\ast_h$
, a contradiction. Note also that Lemma \ref{lemma:ordered} ties the efforts of allies together. It is thus indeed wlog to assume $h'=1$ and $\ell'=0$ since we simply require some players to ``exert high effort". 
We could construct a similar argument if $1$, $0$, or both, were in different components than $i$ and $j$ or when the initial $\ell=0$ and $h=1$. %Since $x^\ast_\ell>x^\ast_h$, suppose wlog that $j\in K_0(g)$. From $\underline{\delta}<\delta^\ast$ we know that $\tau(0-\theta_j)^2\leq\tau[(\theta_j-0)^2+(\theta_j-\bar{t}_j)]$, i.e., the total tolerance cost for $0$ is lower than for $j$. Since $k_0\leq k_j<k_i$ by assumption, one of two things must hold. Either, $\tau[(\underline{t}_i-\theta_i)^2+(\bar{t}_i-\theta_i)^2]>\tau[(0-\theta_j)^2+(\bar{t}_j-\theta_j)^2]$ which would imply a profitable deviation for $i$, since her neighbors exert little effort, or, $\tau[(\underline{t}_i-\theta_i)^2+(\bar{t}_i-\theta_i)^2]\leq\tau[(0-\theta_j)^2+(\bar{t}_j-\theta_j)^2]$, i.e., $j$ pays more for tolerance than $i$. Since $f(\cdot,\cdot,\cdot)$ is strictly concave in its first argument, this would imply $0\not\in K_j(g^\ast)$, a contradiction.
Lemma \ref{lemma:socialization-efforts} follows.
\end{proof}\\

\begin{proof}[Proof of Proposition \ref{proposition:polarization}]
We first establish how the benefits from dispute depend on $\alpha$ and $\beta$ in the parametric model.

\begin{lemma}\label{lemma:beta}
    Take some $\alpha>\alpha'$ and $\beta>\beta'$. Then, $\delta(y,\alpha,\beta)>\delta(y,\alpha,\beta')>\delta(y,\alpha',\beta')$ and $\delta(y,\alpha,\beta)>\delta(y,\alpha',\beta)>\delta(y,\alpha',\beta')$ for all $y\in \{0,1,...,n-2\}$. Moreover, if $\beta=0$, $\delta(y,\alpha,\beta)=0$ for all $y\in\{0,1,...,n-2\}$. %and $\alpha\in\mathbb{R}^{+}_{0}$.
\end{lemma}

\begin{proof}
    Note that $\partial f(x,y,z)/\partial z\geq0$. Moreover, %$\frac{\partial \lambda^\ast_{ij}}{\partial \alpha}\geq0$ and $\frac{\partial \lambda^\ast_{ij}}{\partial \beta}\geq0$. From the envelope theorem, we can infer 
    $\partial f(x,y,z)/\partial \alpha\geq0$ and $\partial f(x,y,z)/\partial \beta\geq0$. The first part of the statement follows directly. If we fix $\beta=0$ and vary $\alpha$, $\partial f(x,y,z)/\partial z=0$. The lemma follows.
\end{proof}\\
    
\noindent Next, we address how dispute intensity and total effort changes in $\alpha$ and $\beta$. There are two cases.%First, take $\underline{\delta}<\underline{\delta'}<\delta^\ast$ and suppose \textit{ad absurdum} $\iota(s^\ast)>\iota(s^{\ast'})$. This implies either there is more dispute under $s^M$ or $\sum_{i\in N}x^\ast_i>\sum_{i\in N}x^{\ast'}_i$. I address the two cases separately.

\noindent \textbf{Case 1:} $\underline{\delta}<\delta^\ast$. By assumption, it is more profitable to have mutual opponents, i.e., $\underline{\delta}$ increases. Since the benefits from dispute are strictly greater, there cannot be fewer disputes in equilibrium. By Lemma \ref{lemma:socialization-efforts} and since the number of disputes increases, total effort increases. This implies an increase on the extensive and intensive margin of dispute intensity. %Moreover, extreme types can form more connections, since returns of having mutual opponents increase. It follows trivially from Lemma \ref{lemma:investment} that those types socialize less.

\noindent \textbf{Case 2:} $\underline{\delta}\geq\delta^\ast$. Society consists of cliques. As $\underline{\delta}$ increases, $\vert C_{\bar{m}}(g^\ast)\vert$ increases, where $\bar{m}$ denotes the largest clique. Clearly, $\bar{m}$ cannot decrease, since cohesion becomes more effective. Since any equilibrium network is ordered (Lemma \ref{lemma:ordered}), there are fewer disputes in society. Lemma \ref{lemma:socialization-efforts} implies total effort is decreasing. Dispute intensity is decreasing on both margins and the statement follows.
\end{proof}\\

\begin{proof}[Proof of Proposition \ref{proposition:information cost}]
We prove the proposition in a series of lemmata.

%\noindent Suppose \textit{ad absurdum} $i$ deletes the link to one of her neighbors, say $j$, and increases her socialization efforts. Hence, $\sum_{h\in K_i(g^\ast)}f'(\lambda_i,\lambda_h,\lambda_{ih})=\sum_{h\in K_i(g^{\ast'})}f'(\lambda'_i,\lambda_h,\lambda_{ih})+f'(\lambda'_i,\lambda'_j,\lambda_{ij})=c$. By Lemma $\ref{lemma:socialization-efforts}$, $x^{\ast'}_i>x^\ast_i$ and $x^{\ast'}_j>x^\ast_j$, however, since $f'(\cdot,\cdot,\cdot)$ is decreasing in its first argument, $\sum_{h\in K_i(g^\ast)}f'(\lambda_i,\lambda_h,\lambda_{ih})-\sum_{h\in K_i(g^{\ast'})}f'(\lambda'_i,\lambda_h,\lambda_{ih})<0$. Hence, $f'(\lambda'_i,\lambda'_j,\lambda_{ij})<0$, a contradiction since $f(\cdot,\cdot,\cdot)$ is increasing in its first argument. Total socialization is thus decreasing. 

\noindent The first Lemma establishes that the network is segregated for a sufficiently high effort cost.

\begin{lemma}
    For sufficiently high $c$, in a dense society, the equilibrium network consists of cliques.
\end{lemma}

\begin{proof}
%To prove the statement, we first show that for $c\to0$ and $c\to\infty$, $g^\ast$ exhibits structural balance. 
The function $f(\lambda_i,\lambda_j,\lambda_{ij})$ is strictly concave in its first argument and $\partial^2f(\lambda_i,\lambda_j,\lambda_{ij})/\partial x_i\partial x_j>0$ if $\lambda_i>\lambda_j$. It is then easy to see that $\lambda^\ast_i\to\lambda^\ast_j$ for all $i,j\in N$ if $c\to\infty$. Suppose \textit{ad absurdum} that there exist $i$, $j$ and $h$, such that $g^\ast_{ij}>0$, $g^\ast_{jh}>0$, and $g^\ast_{ih}=0$, where $i>j>h$ wlog. This implies $\delta(k_i+1)-\delta(k_i)<\tau[(\theta_i-\theta_h)^2-(\theta_i-\underline{t}^\ast_i)^2]$, or $\delta(k_h+1)-\delta(k_h)<\tau[(\theta_h-\theta_i)^2-(\theta_h-\bar{t}^\ast_h)^2]$, or both. Moreover, $\delta(k_j)-\delta(k_j-1)\geq\tau[(\theta_j-\theta_h)^2-(\theta_j-\underline{t'}_j)^2]$, where $t'$ denote deviations for the respective players. There must exist some player $\ell$, such that $g^\ast_{i\ell}=g^\ast_{j\ell}=0$ and a player $\ell'$ such that $g^\ast_{j\ell'}=g^\ast_{h\ell'}=0$. Otherwise, the connections to $i$ and $h$ yield no benefits and $j$ has a profitable deviation in reducing $\bar{t}^\ast_j$ or increasing $\underline{t}^\ast_j$. Since any equilibrium network is ordered, we can focus on the case $\ell>i$ wlog. This implies $\delta(k_i+1)-\delta(k_i)<\tau[(\theta_i-\theta_\ell)^2-(\theta_i-\bar{t}^\ast_i)^2]$. Moreover, $\delta(k_i)-\delta(k_i-1)\geq\tau[(\theta_i-\theta_j)^2-(\theta_i-\bar{t'}_j)^2]$. Since $\lambda_i^\ast\to\lambda_j^\ast$ for all $i,j\in N$, in a dense society, we immediately reach a contradiction that proves the statement.
\end{proof}\\

\noindent The next Lemma establishes that total effort is indeed decreasing in the cost of exerting effort.

\begin{lemma}
    Total effort is decreasing in the effort cost.
\end{lemma}

\begin{proof}
    For each player $i\in N$, the first-order condition dictates $\sum_{h\not\in K_i(g^\ast)}f'(\lambda_i,\lambda_h,\lambda_{ih})=c$. Suppose \textit{ad absurdum} that $x^\ast_i$ increases as $c$ increases. This implies either $i$ is in more disputes or $x^\ast_h$ increases for at least some $h$. The latter case requires that $h$ is in more disputes by a similar argument. We can thus focus on the case where $i$ is in more disputes wlog, say with some player whom we also call $h$. In a segregated society, we directly know total efforts decrease from Lemma \ref{lemma:socialization-efforts} since the number of disputes decreases, as established by the previous lemma. Consider next an overlapping society. As $i$ and $h$ become opponents, there are two effects on their strengths. First, $i$ and $h$ no longer contribute to each others' strengths. Second, each other connection they have obtains a higher weight as by definition of $\rho_{\cdot\cdot}(x,\mathbf{t})$. Since $f(\cdot,\cdot,\cdot)$ is strictly concave in its first argument, for $j\in K_i(g^\ast)$ and $\ell\in K_h(g^\ast)$, $x^\ast_j$ and $x^\ast_\ell$ decrease. Since $f(\cdot,\cdot,\cdot)$ is strictly decreasing and convex in its second argument, $x^\ast_i$ and $x^\ast_h$ must decrease as well. Since we can make a similar argument for each pair $i$ and $h$, total efforts must indeed decrease.
\end{proof}\\

\noindent Next, suppose the cost of exerting effort is high and the network is segregated. An increase in $c$ crowds out efforts, which decreases dispute intensity on the intensive margin. Moreover, %either no existing link is severed or no absent link is added and dispute intensity decreases. Otherwise, players, on average, form more links, since agents are more similar in their strengths and thus coordination on mutual opponents is more beneficial.
the relative benefits from from cohesion increase and the largest clique cannot become smaller. There are fewer disputes in society and dispute intensity decreases on the extensive margin. 

\noindent It remains to be shown that dispute intensity increases for some intermediate $c$. The next lemma addresses this.

\begin{lemma}
    If $\delta^\ast>\underline{\delta}>\tilde{\delta}$, in a dense society, dispute intensity is increasing in $c$. %Moreover, there exist $\Delta_{0}>0$ and $\Delta_1>0$, such that $x^M_i$ is decreasing if $\theta_i\in[0,\Delta_0]$ or $\theta_i\in[1-\Delta_1,1]$. 
\end{lemma}

\begin{proof} Clearly, $\tilde{\delta}<\delta^\ast$ since we would be in the high cost case otherwise. The extensive and intensive margin of dispute intensity go in opposite directions, since increasing the effort cost decreases total efforts (and thus dispute intensity on the intensive margin) but increases the relative effectiveness of cohesion (and thus dispute intensity on the extensive margin). It remains to be shown that there exists some interval $]\tilde{\delta},\delta^\ast]$, such that dispute intensity is increasing in $c$. Take $\underline{\delta}=\delta^\ast+\epsilon$ with $\epsilon\to0$. Hence, a marginal increase in $c$ increases the number of disputes in a dense society since the network moves towards segregation. Dispute intensity increases on the extensive margin. Total efforts decrease. Since the benefits from adding a neighbor with a mutual opponent increase discretely, the increase in dispute intensity on the extensive margin must outweigh the decrease on the intensive margin if the network is sufficiently close to segregation as dispute intensity on the intensive margin is continuous and monotonic in $c$ for the interval $]\tilde{\delta},\delta^\ast]$. The value $\tilde{\delta}$ distinguishes the cases of low and intermediate linking costs, where the decrease in dispute intensity on the intensive margin outweighs the increase in dispute intensity on the extensive margin by construction. The statement follows. 
\end{proof}

\noindent The proposition follows directly from the lemmata.
\end{proof}

\renewcommand{\theequation}{B-\arabic{equation}}
\renewcommand\theproposition{B-\arabic{proposition}}
\renewcommand\thecorollary{B-\arabic{corollary}}
\renewcommand\thetheorem{B-\arabic{theorem}}

\section{Extensions}\label{appendixB}

\begin{proof}[Proof of Proposition \ref{proposition:prize}]
Potential disputes are the zeros in adjacency matrix $g$. To prove the proposition, it is sufficient to show that at least one agent expects positive net benefits from initiating such a potential dispute. Since we allow for self-loops, $\lambda^\ast_i>0$ for all $i\in N$. By Lemma \ref{lemma:ordered}, the bilateral equilibrium network is ordered. Hence, $\lambda^\ast_{ih}>0$, for all $i$ with $K_i(g^\ast)\ne\emptyset$. In a dense society, this is true for all agents. Then, $f(\lambda^\ast_i,\lambda^\ast_h,\lambda_{ih})\geq D$ if $D$ is small enough. %Moreover, if $D$ is small enough relative to $\tau$, $K_i(g^\ast)\ne\emptyset$ for all $i\in N$ in a dense society. %Since we could simply relabel $h$ and $i$, the statement follows.
\end{proof}\\

\noindent We next turn to several extensions. To do so, we require some additional concepts and notation.\\

\noindent \textbf{Network definitions}: There is a path in network $g$ from $i$ to $j$, denoted by $p_{ij}=1$, if either $g_{ij}>0$, or there are $m$ players $j_{1},...,j_{m}$, distinct from $i$ and $j$, with $g_{ij_{1}}>0$, $g_{j_{1}j_{2}}>0,\cdot\cdot\cdot, g_{j_{m}j}>0$. The length of the path, $l(p_{ij})$, is one in the former case and $m+1$ in the latter case. Denote the weight of the path by $w_{ij}=g_{ij_1}g_{j_1j_2}\cdot\cdot\cdot g_{j_{m-1}j}$. If there exists no path from $i$ to $j$, $p_{ij}=0$, and $w_{ij}=0$.\\

\noindent \textbf{Alternative strengths:} One can think of an \textit{adjusted Contest Success Function}, where common connections of two players in dispute do not influence their strengths. In particular, consider $\mu_i=\lambda_i-\sum_{h\in N}g_{ih}\sgn(g_{hj})$. One can use $\mu_i$ instead of $\lambda_i$ in the CSF to obtain similar results. Note also, for the case of $\underline{\delta}>\delta^\ast$, $\mu_i=\lambda_i$. Alternatively, consider a model where higher order neighbors contribute to the strength of players. In particular, denote by $P_i^m=\{j\in N: \exists p_{ij} \text{, with }l(p_{ij})\leq m\}$ the set of players, to whom a path from $i$ of length no more than $m$ exists. Let $w_{ij}=w_{ii_1}g_{i_1i_2}\cdot\cdot\cdot g_{j_mj}$ denote the weight of path $p_{ij}$ of length $l(p_{ij})=m$. Define $\mu_i=\sum_{j\in P^m_i}w_{ij}$. One can use $\mu_i$ instead of $\lambda_i$ in the CSF and obtain similar results. %I highlight instances, where the results do not hold for both specifications.

\noindent\textbf{Flexibility}: Another natural variable of interest is the flexibility of agents, i.e., how easily they can tolerate others. This parameter controls the strength of homophily in our model, i.e., how costly it is to interact with differing ideologies. Through the eyes of our model, some societies may be more accepting of interactions between ideologically dissimilar individuals, so agents incur only small reputation losses when tolerating ideologies far from their own. The next corollary summarizes the result on flexibility.

\begin{corollary}\label{cor:tolerance}
    Dispute intensity is decreasing in the flexibility of agents ($\tau$).
\end{corollary}

\noindent Unsurprisingly, more flexible societies are less polarized. When tolerance is not too costly, agents are willing to tolerate a larger range of ideologies. As already established, more connections lead to lower efforts, thereby reducing the intensity of individual disputes. At the same time, agents are in fewer disputes. This implies dispute intensity also decreases because the number of disputes decreases. However, disputes are a genuine feature of the equilibrium, even in societies with flexible agents. In the contexts we have in mind, there is no value to become an ally with everyone since this would deprive agents of the possibility to campaign against others and derive benefits. 

\end{appendices}

\end{document}